\documentclass[prc,showpacs,showkeys,superscriptaddress,floatfix,twocolumn]{revtex4}
\usepackage[bookmarks,dvips,pdfhighlight=/O,pdfstartview=FitH]{hyperref}
\usepackage{graphicx,color,epstopdf,mathrsfs,amsmath,amssymb,amsthm,amsopn,mathbbol,bm}
\usepackage{amsfonts}
\usepackage{amsbsy}
\def\lsim{\mathrel{\rlap{
\lower4pt\hbox{\hskip-3pt$\sim$}}
    \raise1pt\hbox{$<$}}}     
\def\gsim{\mathrel{\rlap{
\lower4pt\hbox{\hskip-3pt$\sim$}}
    \raise1pt\hbox{$>$}}}     
\def\scr#1{\mbox{\scriptsize #1}}
\usepackage{graphicx}
\begin{document}
\title{Relativistic Heavy-Ion Collisions within 3-Fluid 
Hydrodynamics\footnote{http://theory.gsi.de/$\sim$mfd/}%
:\\
Hadronic Scenario}
\author{Yu.B.~Ivanov}\thanks{e-mail: Y.Ivanov@gsi.de}
\affiliation{Gesellschaft
 f\"ur Schwerionenforschung mbH, Planckstr. 1,
D-64291 Darmstadt, Germany}
\affiliation{Kurchatov Institute, Kurchatov
sq. 1, Moscow 123182, Russia}
\author{V.N.~Russkikh}\thanks{e-mail: russ@ru.net}
\affiliation{Gesellschaft
 f\"ur Schwerionenforschung mbH, Planckstr. 1,
D-64291 Darmstadt, Germany}
\affiliation{Kurchatov Institute, Kurchatov
sq. 1, Moscow 123182, Russia}
\author{V.D.~Toneev}\thanks{e-mail: V.Toneev@gsi.de}
\affiliation{Gesellschaft
 f\"ur Schwerionenforschung mbH, Planckstr. 1,
D-64291 Darmstadt, Germany}
\affiliation{Joint Institute for Nuclear Research,
141980 Dubna, Moscow Region, Russia}
\begin{abstract}
A 3-fluid hydrodynamic  model for simulating relativistic heavy-ion
collisions is introduced. Alongside with two baryon-rich fluids, the
new model considers time-delayed evolution of a third, baryon-free
(i.e. with zero net baryonic charge) fluid of newly produced particles. 
Its evolution is
delayed due to a formation time $\tau$, during which the baryon-free fluid
neither thermalizes nor interacts with the baryon-rich fluids. 
After the formation it starts to interact with the
baryon-rich fluids and quickly gets thermalized. 
Within this
model with pure  hadronic equation of state, a
systematic analysis of various observables  at incident energies
between few and about 160$A$ GeV has been done as well as
comparison with results of transport models. We have succeeded to reasonably
reproduce a great 
body of experimental data in the incident energy range of
$E_{\scr{lab}}\simeq$ (1--160)$A$ GeV. The list includes proton
and pion rapidity distributions, proton transverse-mass spectra,
rapidity distributions of $\Lambda$ and  $\bar{\Lambda}$ hyperons, 
elliptic flow of protons and pions (with the exception of
proton $v_2$ at 40$A$ GeV), multiplicities of pions, positive
kaons, $\phi$ mesons, 
hyperons and antihyperons, including multi-strange particles. This agreement is
achieved on the expense of substantial enhancement of the
interflow friction as compared to that estimated proceeding from
hadronic free cross sections. 
 However, we have also found
out certain problems. The calculated yield of $K^-$ is approximately
by a factor of 1.5 higher than that in the experiment. 
We have also failed to describe directed transverse
flow of protons and pion at $E_{\scr{lab}}\geq$ 40$A$ GeV.
This failure apparently indicates that the used EoS is
too hard and thereby leaves room for a phase transition. 
\end{abstract}
%
\maketitle

\section{Introduction}

Relativistic nucleus-nucleus collisions at incident energies
$E_{\scr{lab}}\simeq$ (10--40)$A$ GeV presently attracts
special attention, since the highest baryon densities
\cite{Friman98,CB} and highest relative strangeness \cite{BCOR02} at
moderate temperatures are expected in this incident energy range.
The interest to this energy range has been recently also revived
in connection with new experimental results from low-energy scanning
SPS program \cite{SPS} and the project of the new accelerator
facility SIS300 at GSI \cite{SIS200}. This domain of the nuclear
phase diagram is less explored both experimentally and
theoretically as compared to the corresponding extremes, i.e. cold
compressed matter and baryon-free hot matter. Moreover, a critical
point of the QCD phase diagram may occur to be accessible in these
reactions \cite{Fodor01,Stephanov99}. There are already available
experimental data \cite{SPS} pointing out that something
interesting really happens in this energy range. In order to
draw information on properties of hot and compressed nuclear
matter from available experimental data, we need better
understanding of dynamics of nucleus-nucleus collisions under
investigation.

A direct way to address thermodynamic properties of the matter
produced in these reactions consists in application of
hydrodynamic simulations to nuclear collisions. However, finite
nuclear stopping power, revealing itself at high incident
energies, makes the collision dynamics non-equilibrium and
prevents us from application of conventional hydrodynamics
especially at the initial stage of the reaction. Since the
resulting  non-equilibrium is quite strong, introduction of
viscosity and thermal conductivity does not help to overcome this
difficulty, because by definition they are suitable for weak
non-equilibrium. A possible way out is taking advantage of a
multi-fluid approximation to heavy-ion collisions, pioneered by
Los-Alamos group \cite{Amsden,Clare} and further developed at the
Kurchatov Institute \cite{MRS88,MRS91,INNTS}, Frankfurt
\cite{Kat93,Brac97,Brac00a} and GSI \cite{gsi94,gsi91}. The last
development \cite{gsi94,gsi91}, i.e. the mean-field dynamics in
the multi-fluid background, concerns mostly moderate energies
around 1 GeV/nucleon and therefore will not be discussed here.

The first extension of the original 2-fluid model
\cite{Amsden,Clare}, the two-fluid hydrodynamics with
free-streaming radiation of pions, was advanced in~\cite{MRS88}.
The initial stage of heavy-ion collisions definitely is a highly
non-equilibrium process. Within the hydrodynamic approach this
non-equilibrium is simulated by means of a 2-fluid approximation,
which takes care of the finite stopping power of nuclear matter
\cite{IMS85,IS85}, and simultaneously describes the entropy
generation at the initial stage. The radiated pions form a
baryon-free matter in the mid-rapidity region, while two
baryon-rich fluids simulate the propagation of leading particles.
The pions are the most abundant species of the baryon-free matter
which may contain any hadronic, including baryon-antibaryon pairs,
and/or quark-gluon species.

First applications of the 2-fluid model with direct pion radiation
\cite{MRS91,INNTS} were quite successful for describing heavy-ion 
collisions in the wide range of incident energies, from SIS to SPS. In
these 3D hydrodynamic simulations the  
whole process of the reaction was described, i.e. the evolution
from the formation of a hot and dense nuclear system to its
subsequent decay. 
This is in distinction to numerous other
simulations, which treat only the expansion stage of a fireball
formed in the course of the reaction, while the initial state of
this dense and hot nuclear system is constructed 
from either
kinetic simulations \cite{kin-hydro}
or more general albeit model-dependent
assumptions (e.g. see \cite{1-hydro,HS95}).

However, the approximation of free-streaming pions, produced in
the mid-rapidity region, was still irritating from the theoretical
point of view, in particular, because the relative momenta of the
produced pions and the leading baryons are in the range of the
$\Delta$ resonance for the incident energies considered. This
would imply that the interaction between the produced pions and
the baryon-rich fluids should be strong. The free-streaming
assumption relies on a long formation time of  produced pions.
Indeed, the proper time for the  formation of the produced
particles   is commonly assumed to be of the order  1 fm/c in the
comoving frame. Since the main part of the produced pions is quite
relativistic at high incident energies, their formation time
should be long enough in the reference frame of calculation to
prevent them from interacting with the baryon-rich fluids.
However, this argument is qualitative rather than quantitative,
and hence requires further verification. The first attempt to do
this was undertaken by the Frankfurt group \cite{Kat93}, which
started to explore an opposite extreme. They assumed that the
produced pions immediately thermalize, forming a baryon-free fluid
(or a ``fireball'' fluid, in terms of \cite{Kat93}), and interact
with the baryon-rich fluids. No formation time was allowed, and
the strength of the corresponding interaction was guessed rather
than microscopically estimated. This opposite extreme, referred to
as a (2+1)-fluid model and being not quite justified either,
yielded results substantially different from those of the
free-streaming approximation. This was one of the reasons why in
subsequent applications the Frankfurt group neglected the
interaction between baryon-free and baryon-rich fluids while
keeping the produced pions thermalized \cite{Brac97,Brac00a}, thus
effectively restoring the free-streaming approximation. However,
the assumed immediate thermalization of the fireball fluid
together with the lack of interaction with baryon-rich fluids
still was not a consistent approximation.

In the present paper we would like to introduce an extension of
the multi-fluid approach for simulating heavy-ion collisions, i.e.
a 3-fluid model with formation time. This model is a
straightforward extension of the 2-fluid model with radiation of
direct pions \cite{MRS88,MRS91,INNTS} and (2+1)-fluid model
\cite{Kat93,Brac97,Brac00a}. We extend the above models in such a
way that the created baryon-free fluid (which we call a
``fireball'' fluid, according to the Frankfurt group) is treated
on equal footing with the baryon-rich ones. 
This implies that we
allow a certain formation time for the fireball fluid, during
which the matter of the fluid propagates without interactions. 
Moreover, we assume that the fireball matter gets quickly thermalized
after its formation. The latter approximation is an enforced one, since
we deal with the hydrodynamics rather than with kinetics. 
The interaction between fireball and baryon-rich
fluids is estimated based on elementary cross sections.
A brief
account of this model has been already reported in Ref. \cite{3f-yaf}.

The formation time ($\tau$) is a conventional tool of the hadronic physics, 
which is  associated with a finite time of string formation. It is 
incorporated in kinetic transport models such as UrQMD \cite{Bass98} 
and HSD \cite{Cassing99}. 
In dense medium the separate strings can already interact, this is the 
reason of introduction of junctions and formation of color ropes. 
In fact, this string interaction is a method of extending the treatment
beyond the approximation of binary collisions which is inherent to
the conventional Boltzmann equation. This interaction does not 
invalidate the concept of formation time but rather extends it to 
the case of multi-particle collisions. The concept of strings becomes 
irrelevant in the deconfined state of matter. Here there are two points 
of view: (i) quarks and gluons are instantly produced accordingly to 
the QCD perturbation theory, or (ii) they are mediated by a coherent 
color field \cite{McLerran94,Kapusta02}, i.e. first the coherent 
color field is produced which subsequently decays into incoherent 
fluctuations -- quarks and gluons. The first mechanism implies that 
$\tau$=0, while the second one still makes room for a 
finite $\tau$. Therefore, the fitted value of $\tau$ may help to 
distinguish between these two mechanisms of the QGP production.

The developed code allows calculations with various equations of
states (EoS), which enter as a separate block of the code. We have
started with the simplest, purely hadronic EoS which involves only
a simple density dependent mean field providing saturation of cold
nuclear matter at normal nuclear density $n_0=$ 0.15 fm$^{-3}$ and
with the proper binding energy -16 MeV. This EoS is a natural
reference point for any other more elaborate EoS. Much to our
surprise, this trivial EoS  turned out to be able to reasonably
reproduce a great body of experimental data. 
Taking advantage of the modern computers, substantial work has
been also done on improvement of numerics of the model. 
These results are reported in this paper.

\section{3-Fluid Hydrodynamic Model with Delayed Formation}

The derivation of equations of the 3-fluid model 
presented below is in fact only an illustration of physical 
assumptions which this model is based on. Indeed, the Boltzmann 
equation, from which this derivation starts, strictly speaking is 
inapplicable to a dense and strongly interacting matter. Therefore, 
the applicability of the 3-fluid model to heavy-ion collisions is 
certainly an assumption which should be verified in comparison 
with experimental data. However, this is a general 
situation with dense systems even in simpler cases. 
The basic reason for introduction of the 3-fluid approximaion is 
simulation of the finite stopping power which is important at the 
formation stage of the initial hot and dense blob of nuclear matter. 
In this sense, it is an alternative to constructing this initial blob
by means of either various kind kinetic transport models
\cite{kin-hydro} or model assumptions \cite{1-hydro,HS95}. 

\subsection{Basic Formulation}

Unlike the conventional hydrodynamics, where local instantaneous
stopping of projectile and target matter is assumed, a specific
feature of the dynamic 3-fluid description is a finite stopping
power resulting in a counter-streaming regime of leading
baryon-rich matter.  Experimental rapidity distributions in
nucleus--nucleus  collisions support this counter-streaming
behavior, which can be observed for incident energies  between few
and 200$A$ GeV. The basic idea of a 3-fluid approximation to
heavy-ion collisions \cite{MRS88,I87} is that at each space-time
point $x=(t,{\bf x})$ the generally nonequilibrium 
distribution function of baryon-rich
matter, $f_{\scr{br}}(x,p)$, can be represented as a sum of two
distinct contributions
\begin{eqnarray}
\label{t1} f_{\scr{br}}(x,p)=f_{\scr p}(x,p)+f_{\scr t}(x,p),
\end{eqnarray}
initially associated with constituent nucleons of the projectile
(p) and target (t) nuclei. In addition, newly produced particles,
populating the mid-rapidity region, are associated with a fireball
(f) fluid described by the  distribution function $f_{\scr
f}(x,p)$. 
Therefore, the 3-fluid approximation is a minimal way to 
simulate the finite stopping power at high incident energies.
Note that both the baryon-rich and fireball fluids may
consist of any type of hadrons  and/or partons (quarks and
gluons), rather than only nucleons and pions. However, here and
below we suppress the species label at the distribution functions
for the sake of transparency of the equations.

With the above-introduced distribution functions $f_\alpha$
($\alpha=$p, t, f), the coupled set of relativistic Boltzmann
equations looks as follows:
\begin{eqnarray}
p_\mu\partial^\mu_x f_{\scr p} (x,p) &=& C_{\scr p} (f_{\scr
p},f_{\scr t})+C_{\scr p} (f_{\scr p},f_{\scr f}),
   \label{t2}
\\
p_\mu\partial^\mu_x f_{\scr t} (x,p) &=& C_{\scr t} (f_{\scr
p},f_{\scr t})+C_{\scr t} (f_{\scr t},f_{\scr f}),
   \label{t3}
\\
p_\mu\partial^\mu_x f_{\scr f} (x,p) &=& C_{\scr f} (f_{\scr
p},f_{\scr t}) +C_{\scr f} (f_{\scr p},f_{\scr f})+C_{\scr f}
(f_{\scr t},f_{\scr f}),
   \label{t4}
\end{eqnarray}
where $C_\alpha$ denote collision terms between the constituents
of the three fluids. We have omitted intra-fluid collision terms,
like $C_{\scr p} (f_{\scr p},f_{\scr p})$, since below they will
be canceled anyway. The displayed inter-fluid collision terms have
a clear physical meaning: $C_{\scr p/t} (f_{\scr p},f_{\scr t})$,
$C_{\scr p/t} (f_{\scr p/t},f_{\scr f})$, and $C_{\scr f} (f_{\scr
p/t},f_{\scr f})$ give rise to friction between p-, t- and
f-fluids, and the term $C_{\scr f} (f_{\scr p},f_{\scr t})$ takes care of
particle production in the mid-rapidity region. Note that up to
now we have done no approximation, except for hiding intra-fluid
collision terms.

Let us proceed to approximations which justify the term ``fluids''
having been used already. We assume that constituents within each
fluid are  locally equilibrated, both thermodynamically and
chemically, i.e. that $f_\alpha$ are equilibrium distributions. 
In particular, this implies that the intra-fluid
collision terms are indeed zero. This assumption relies on the
fact that  intra-fluid collisions are much more efficient in
driving a system to equilibrium than  inter-fluid interactions. As
applied to the fireball fluid, this assumption requires some
additional comments, related to the concept of a finite formation
time. During the proper formation time $\tau$ after production,
the fireball fluid propagates  freely, interacting neither  with
itself nor with the baryon-rich fluids. 
After this time interval, the fireball matter starts to interact
with both itself and the baryon-rich fluids and, as a result, locally
thermalizes. 
Being heated up,
 these three fluids may contain not only hadronic and but also
 deconfined  quark-gluon species, depending on the EoS used.

The above assumption suggests that interaction between different
fluids should be treated dynamically. To obtain the required
dynamic equations, we first integrate the kinetic
Eqs.~(\ref{t2})--(\ref{t4}) over momentum and sum over particle
species with weight of baryon charge. This way we arrive to
equations of the baryon charge conservation
   \begin{eqnarray}
   \label{eq8}
   \partial_{\mu} J_{\alpha}^{\mu} (x) &=& 0,
   \end{eqnarray}
for $\alpha=$p and t, where
$J_{\alpha}^{\mu}=n_{\alpha}u_{\alpha}^{\mu}$ is the baryon
current defined in terms of baryon density $n_{\alpha}$ and
 hydrodynamic 4-velocity $u_{\alpha}^{\mu}$ normalized as
$u_{\alpha\mu}u_{\alpha}^{\mu}=1$. Eq.~(\ref{eq8}) implies that
there is no baryon-charge exchange between p- and t-fluids, as
well as that the baryon current of the fireball fluid is
identically zero, $J_{\scr f}^{\mu}=0$. Integrating kinetic
Eqs.~(\ref{t2})--(\ref{t4}) over momentum with weight of
4-momentum $p^\nu$ and  summing over all particle species, we
arrive at equations of the energy--momentum exchange for
energy--momentum tensors $T^{\mu\nu}_\alpha$ of the 
fluids
   \begin{eqnarray}
   \partial_{\mu} T^{\mu\nu}_{\scr p} (x) &=&
-F_{\scr p}^\nu (x) + F_{\scr{fp}}^\nu (x),
   \label{eq8p}
\\
   \partial_{\mu} T^{\mu\nu}_{\scr t} (x) &=&
-F_{\scr t}^\nu (x) + F_{\scr{ft}}^\nu (x),
   \label{eq8t}
\\
   \partial_{\mu} T^{\mu\nu}_{\scr f} (x) &=&
F_{\scr p}^\nu (x) + F_{\scr t}^\nu (x) - F_{\scr{fp}}^\nu (x) -
F_{\scr{ft}}^\nu (x),
   \label{eq8f}
   \end{eqnarray}
where the $F^\nu_\alpha$ are friction forces originating from
inter-fluid collision terms in the kinetic
Eqs.~(\ref{t2})--(\ref{t4}). $F_{\scr p}^\nu$ and $F_{\scr t}^\nu$ in
Eqs.~(\ref{eq8p})--(\ref{eq8t}) describe energy--momentum loss of
baryon-rich fluids due to their mutual friction. A part of this
loss $|F_{\scr p}^\nu - F_{\scr t}^\nu|$ is transformed into
thermal excitation of these fluids, while another part $(F_{\scr
p}^\nu + F_{\scr t}^\nu)$ gives rise to particle production into
the fireball fluid (see Eq.~(\ref{eq8f})). $F_{\scr{fp}}^\nu$ and
$F_{\scr{ft}}^\nu$ are associated with friction of the fireball
fluid with the p- and t-fluids, respectively. Note that
Eqs.~(\ref{eq8p})--(\ref{eq8f})  satisfy the  total
energy--momentum conservation
\begin{eqnarray}
\partial_\mu (T^{\mu\nu}_{\scr p} +
T^{\mu\nu}_{\scr t} + T^{\mu\nu}_{\scr f}) = 0. \label{eq10}
\end{eqnarray}

As described above, the energy--momentum tensors of the
baryon-rich fluids ($\alpha=$p and t) take the conventional
hydrodynamic form
   \begin{eqnarray}
T^{\mu\nu}_\alpha= (\varepsilon_\alpha + P_\alpha) \
u_{\alpha}^{\mu} \ u_{\alpha}^{\nu} -g^{\mu\nu} P_\alpha
   \label{eq11}
   \end{eqnarray}
in terms of the proper energy density, $\varepsilon_\alpha$, and
pressure, $P_\alpha$. 
For the fireball, however, only the energy--momentum tensor of the
formed matter is of
practical interest for us. Since we treat the fireball matter as a
fluid, we have nothing to do but assume that it is formed already
thermalized. This assumption implies that its thermalization time is
essentially shorter than its formation time. 
Moreover, this assumption is in the spirit of 
other assumptions made: we distinguish the main nonequilibrium 
associated with finite stopping power and consider all the rest of
dynamics within local equilibrium. 
Thus, only the formed
(and by assumption thermalized) part of the energy--momentum tensor is
described by this hydrodynamic form
   \begin{eqnarray}
T^{\scr{(eq)}\mu\nu}_{\scr f}= (\varepsilon_{\scr f} + P_{\scr f})
\ u_{\scr f}^{\mu} \ u_{\scr f}^{\nu} -g^{\mu\nu} P_{\scr f}.
   \label{eq12}
   \end{eqnarray}
Its evolution is defined by an Euler equation with a retarded
source term
   \begin{eqnarray}
\hspace*{-8mm}
&&   \partial_{\mu} T^{\scr{(eq)}\mu\nu}_{\scr f} (x) 
=- F_{\scr{fp}}^\nu (x) - F_{\scr{ft}}^\nu (x)
\cr
\hspace*{-8mm}
&&+
\int d^4 x' \delta^4 \left(\vphantom{I^I_I} x - x' - U_F
(x')\tau\right)
 \left[F_{\scr p}^\nu (x') + F_{\scr t}^\nu (x')\right],
   \label{eq13}
   \end{eqnarray}
where $\tau$ is the formation time, and
   \begin{eqnarray}
   \label{eq14}
U^\nu_F (x')=
\frac{u_{\scr p}^{\nu}(x')+u_{\scr t}^{\nu}(x')}%
{|u_{\scr p}(x')+u_{\scr t}(x')|}
   \end{eqnarray}
is a free-propagating 4-velocity of the produced fireball 
matter, which is evidently a time-like 4-vector. 
In fact, this is the velocity at the moment of production of the
fireball matter. According to Eq.~(\ref{eq13}), the energy and
momentum of this matter appear as a source in the Euler equation
only later, i.e. after the time interval $U_F^0\tau$ upon the
production, and in 
different space point ${\bf x}' - {\bf U}_F (x') \ \tau$, as
compared to the production point ${\bf x}'$. From the first
glance, one can immediately simplify the r.h.s. of Eq.
(\ref{eq13}) by performing integration with the $\delta$-function.
However, this integration is not so straightforward, since the
expression under the $\delta$-function, $x - x' - U_F (x')\tau=0$,
may have more than one solution with respect to $x'$. The latter
would mean that the matter produced in several different
space-time points $x'$ 
is simultaneously formed in the same space-time point $x$. 
This is possible due to the nonlinearity of
the hydrodynamic equations.

The above discussed free-propagating fireball matter is, of course, 
a certain approximation for treating free-streaming particles, which
this matter consists of. These produced free-streaming particles are
characterized by a distribution function, $f(p)$, in the momentum
space, which is determined by their production cross sections, e.g. 
$p^0 d\sigma_{NN\to \pi X}/d^3 p$. In particular, this means that
particles get formed at different time instants 
$\tau_{\scr{particle}} \gamma_{\scr{particle}}$, depending on the
particle velocity ($\gamma_{\scr{particle}}$ is the gamma factor of
the particle). This particle formation is governed by a particle
formation time $\tau_{\scr{particle}}$ which differs from our $\tau$
we use in the hydrodynamic formulation. Our $\tau$ has a meaning of
the particle formation time $\tau_{\scr{particle}}$ averaged over the
distribution $f(p)$: 
\begin{eqnarray}
   \label{tau-part}
\tau\;\gamma_{\scr{f}} = \int \frac{d^3 p}{p^0} 
\tau_{\scr{particle}} \;\gamma_{\scr{particle}} f(p)\left/
\int \frac{d^3 p}{p^0} f(p)\right. , 
\end{eqnarray}
where $\gamma_{\scr{f}}$ is the gamma factor of the unformed fireball
matter. 
Eq. (\ref{tau-part}) shows that $\tau$ is always longer than 
$\tau_{\scr{particle}}$. In particular, if the fireball matter consists 
of pions, which have approximately thermal distribution with temperature 
$\approx 100$ MeV, then $\tau \approx 2\tau_{\scr{particle}}$. 

The free-streaming
particles are not only formed at different time instants but also in
different space points because of the same $f(p)$ distribution. Thus,
another approximation we have done consists in neglecting  this
spacial spread. We could take it into account by changing the
$\delta$-function in the r.h.s. of Eq. (\ref{eq13}) to a smooth
distribution of formed particles. This would result in spatial
smearing of formation of the fireball fluid. However, at the same time
it would make the model formulation more complex, since this smooth
distribution of formed particles should depend on the initial $f(p)$
distribution and, hence, on the cross sections, like  
$p^0 d\sigma_{NN\to \pi X}/d^3 p$. Moreover, it would be another
source of uncertainty, since the $NN\to \pi X$ process is not the
only relevant one. Therefore, for the first estimate of effects of the
formation time we avoid these complications.

The residual part of $T^{\mu\nu}_{\scr f}$ 
(the free-propagating one) 
is defined as
   \begin{eqnarray}
T^{\scr{(fp)}\mu\nu}_{\scr f}= 
T^{\mu\nu}_{\scr f}-T^{\scr{(eq)}\mu\nu}_{\scr f}.
   \label{eq15}
   \end{eqnarray}
The equation for $T^{\scr{(fp)}\mu\nu}_{\scr f}$ can be easily
obtained by taking the difference between Eqs.~(\ref{eq8f}) and
(\ref{eq13}). If all  the fireball matter turns out to be formed
before freeze-out, then this equation is not needed. Thus, the
3-fluid model introduced here contains both the original  2-fluid
model with pion radiation \cite{MRS88,MRS91} and the (2+1)-fluid
model \cite{Kat93,Brac97,Brac00a} as limiting cases for $\tau
\rightarrow \infty$ and $\tau=0$, respectively.

\subsection{Friction between Baryon-Rich Fluids and their Unification}
\label{Friction between Baryon-Rich Fluid}

The nucleon--nucleon cross sections at high energies are strongly
forward--backward peaked. This fact, which originally served as
justification for subdividing baryonic matter into target and
projectile fluids,
was used in \cite{IMS85} to estimate the friction forces, $F_{\scr
p}^\nu$ and $F_{\scr t}^\nu$, proceeding from only $NN$ elastic
scattering. Later these friction forces were calculated
\cite{Sat90} based on (both elastic and inelastic) experimental
inclusive proton--proton cross sections. In the present
calculations we use the following form of the projectile--target
friction
 \begin{eqnarray}
 F_{\alpha}^\nu=\vartheta^2\rho^{\xi}_{\scr p} \rho^{\xi}_{\scr t}
\left[\left(u_{\alpha}^{\nu}-u_{\bar{\alpha}}^{\nu}\right)D_P+
\left(u_{\scr p}^{\nu}+u_{\scr t}^{\nu}\right)D_E\right],
\label{eq16}
\end{eqnarray}
$\alpha=$p or t, $\bar{\mbox{p}}=$t and  $\bar{\mbox{t}}=$p.
Here, $\rho^{\xi}_\alpha$  denotes a kind of "scalar
densities" of the p- and t-fluids (see details below),
\begin{eqnarray}
D_{P/E} = m_N \ V_{\scr{rel}}^{\scr{pt}} \ \sigma_{P/E}
 (s_{\scr{pt}}),
\label{eq17}
\end{eqnarray}
where $m_N$ is the nucleon mass, $s_{\scr{pt}}=m_N^2 \left(u_{\scr
p}^{\nu}+u_{\scr t}^{\nu}\right)^2$ is the mean invariant energy
squared of two colliding nucleons from the p- and t-fluids,
\begin{eqnarray}
\label{Vpt} 
V_{\scr{rel}}^{\scr{pt}}=
[s_{\scr{pt}}(s_{\scr{pt}}-4m_N^2)]^{1/2}/2m_N^2
\end{eqnarray}
is the mean relative velocity of the p- and t-fluids, and
$\sigma_{P/E}(s_{\scr{pt}})$ are determined in terms of
nucleon-nucleon cross sections integrated with certain weights
(see \cite{MRS88,MRS91,Sat90} for details):  
\begin{eqnarray}
\label{sigma_P} 
\hspace*{-5mm}
\sigma_{P}(s_{\scr{pt}}) &\!\!\!=\!\!\!& 
\int_{\theta_{\scr{cm}}<\pi/2} 
d\sigma_{NN\to NX} \left(1-\cos\theta_{\scr{cm}}
\frac{p_{\scr{out}}}{p_{\scr{in}}}\right),
\\
\label{sigma_E} 
\hspace*{-5mm}
\sigma_{E}(s_{\scr{pt}}) &\!\!\!=\!\!\!& 
\int_{\theta_{\scr{cm}}<\pi/2} 
d\sigma_{NN\to NX} \left(1-
\frac{E_{\scr{out}}}{E_{\scr{in}}}\right). 
\end{eqnarray}
Here the integration is restricted to the forward hemisphere 
($\theta_{\scr{cm}}<\pi/2$) of the 
center-of-mass scattering angles $\theta_{\scr{cm}}$, 
$p_{\scr{in}}=(s_{\scr{pt}}/4-m_N^2)^{1/2}$ and 
$E_{\scr{in}}=s_{\scr{pt}}^{1/2}/2$ are the in-coming momentum and
energy of the nucleon before the scattering in the NN c.m. frame,
respectively, and $p_{\scr{out}}$ and $E_{\scr{out}}$ are the
corresponding out-coming quantities. $\sigma_{P}(s_{\scr{pt}})$
is a kind of a transport cross section, which is nonzero at any physical
$s_{\scr{pt}}$, as it is seen from Eq. (\ref{sigma_P}). 
At the same time, the $\sigma_{E}(s_{\scr{pt}})$ quantity, which is
responsible for the fireball production, is
zero for $s_{\scr{pt}}$ below the inelastic threshold. In
particular, the latter feature makes $D_E=0$, and hence the friction
$F_{\alpha}=0$, at vanishing relative velocity $u_{\scr{p}}-u_{\scr{t}}=0$. 
The overall
$\vartheta^2$ factor in Eq. (\ref{eq16}) is associated with
unification of p- and t-fluid into a single one (see Eq.
(\ref{alpha})), when their relative velocity gets small enough.

The above friction (\ref{eq16}) is a certain extension of that
derived in \cite{Sat90} and used in
\cite{MRS88,MRS91,INNTS,Kat93,Brac97}. This original derivation was
performed under assumption that baryon-rich fluids consist of only
nucleons, or baryons, assuming that baryon--baryon cross sections
are similar to proton--proton ones. Therefore, in the original
expression for $F_{\alpha}^\nu$
\cite{MRS88,MRS91,Kat93,INNTS,Brac97,Sat90} modified scalar densities of
the p- and t-fluids are substituted by corresponding baryon
densities (with $\xi_h=\xi_q=1$ and  $\vartheta^2=1$, see below).
However, this original calculation is incomplete, because it does not take
into account~:  
\\
(i) various mesonic (and maybe, quark and gluonic) species
produced in the collision
\\
(ii) possible multiparticle interactions which are quite probable
in the dense medium,
\\
(iii) possible medium modifications of cross sections and effective
masses, and 
\\
(iv) quark and gluon interactions, if the system happens to
undergo the phase transition to the quark-gluon phase.
\\
(v) Moreover, even experimental cross sections between hadronic
species are only poorly known. As for the nonperturbative
quark-gluon phase, this is the range of pure speculations.
\\
In view of theses uncertainties, it is reasonable to make
provision for tuning the above friction. For this purpose we
introduced tuning factors $\xi (s_{pt})$ in the scalar densities
of the p- and t-fluids
\begin{eqnarray}
\label{ro_xi} 
\rho_{\alpha}^\xi(s_{pt}) &=&
\left(\rho_{\alpha}^{bar.}+\frac{2}{3}\rho_{\alpha}^{mes.}\right)\xi_h(s_{pt})
\cr
&+&
\frac{1}{3}\left(\rho_{\alpha}^{q}+\rho_{\alpha}^{g}\right)\xi_q(s_{pt}),
\end{eqnarray}
where 
$\rho_{\alpha}^{bar.}$, $\rho_{\alpha}^{mes.}$,
$\rho_{\alpha}^{q}$ and $\rho_{\alpha}^{g}$ are scalar densities
of all baryons, all mesons, quarks and gluons, respectively, 
defined in the conventional way, i.e.
\begin{eqnarray}
\label{ro_a} \rho_a (x)= m_a \int \frac{d^3 p}{p_0} f_a (x,p)
\end{eqnarray}
for the $a$ species of mass $m_a$ which are
$\rho_{\alpha}(T(x),\mu_B(x))$ in the thermodynamic limit. 
These scalar densities (but of course, only nucleon ones) 
result from  the original derivation of Ref. \cite{Sat90}. 
We just extended the 
recipe of Ref. \cite{Sat90} to other species. In fact, in the 
nonrelativistic case the scalar density is identical to the usual 
particle-number density, since the mass $m_a$ in Eq. (\ref{ro_a}) 
is canceled by $p_0$ in the denominator. 
Moreover, in the relativistic case the scalar density is a natural result, 
since it is a Lorentz invariant contrary to the particle-number density, 
which is the 4th component of a 4-vector. However, in the relativistic 
case we run into trouble with gluons and quarks. If one uses current
masses for gluons and quarks, the corresponding scalar densities are
either identically zero (for gluons) or negligibly small (for $u$ and 
$d$ quarks). To overcome this 
problem we imply that either thermal effective masses or constituent masses, 
depending on particular model, should be used in the scalar densities,
or they should be calculated beyond the quasiparticle approximation
(e.g. on the lattice).

Factors like 2/3 and 1/3 in Eq.
(\ref{ro_xi}) take into account the assumed scaling of cross sections in
accordance with the naive valence-quark counting. Note that the
original free cross section in (\ref{eq16}) was estimated for the 
proton-proton pair. In Eq. (\ref{ro_xi}) different tuning factors
are introduced for hadronic and quark-gluon phase: $\xi_h$ and
$\xi_q$,  respectively. If the system occurs in pure hadronic
phase, $\xi_h^2$ directly scales the cross sections
$\sigma_{P/E}$. If the system happens to be in a mixed phase, like
in the cross-over phase transition, both $\xi_h$ and $\xi_q$
participate in the cross-section scaling. Note that originally the
scalar densities depend only on $x$, cf. Eq. (\ref{ro_xi}). The
introduced $s$-dependent tuning factors make the tuned scalar
densities $\rho_{\alpha}^\xi$ also $s$-dependent. In general, $\xi_h$
and $\xi_q$ could be also assumed to be temperature dependent, since
the $\sigma_{P/E}$ are obtained by averaging over certain thermal
distributions. Nevertheless, we avoid to do this in order to keep the
number of ``fitting degrees of freedom'' as low as possible. 

When the baryon-rich fluids get decelerated enough, it is
reasonable to unify them into a single fluid. This limit is not
automatically present in the model formulation and therefore
should be stipulated\footnote{Note that in the mean-field
multi-fluid dynamics \cite{gsi94,gsi91} this problem is
overcome.}. For this purpose we introduce an auxiliary function
\begin{eqnarray}
\label{alpha} \vartheta = 1 -
\exp[-(V_{\scr{rel}}^{\scr{pt}}/\Delta V)^4]
\end{eqnarray}
where $V_{\scr{rel}}^{\scr{pt}}$ is the mean relative velocity of
the p- and t-fluids of Eq. (\ref{Vpt}), and $\Delta V$ is the
characteristic velocity of particles inside the flow (the Fermi
velocity of nucleons in cold matter or the thermal velocity of
particles in hot matter)
\begin{eqnarray}
\label{DV} 
\left[1-(\Delta V)^2\right]^{-1/2} -1 =
\max_{p,t}\left\{ \frac{\varepsilon_F}{m_N}, \frac{3 T}{2m_N}\right\},
\end{eqnarray}
where $\varepsilon_F$ is the Fermi energy corresponding to the
baryon density of the fluid, $T$ is temperature of the fluid, and
the maximum value is searched over quantities corresponding to the two
overlapped (p- and t-) fluids. Thus, at $V_{\scr{rel}}^{\scr{pt}}
\gg \Delta V$, $\vartheta=1$ and the purely two-fluid regime is
realized, while at $V_{\scr{rel}}^{\scr{pt}} \ll \Delta V$, 
$\vartheta=0$  and
the one-fluid regime takes place, which implies complete
equilibration of the overlapping fluids. During the
numeric simulation the two-fluid and one-fluid solutions are mixed
in proportion $\vartheta$ and $1-\vartheta$, respectively, thus
providing a smooth approach to the one-fluid limit. 
This interpolation procedure concerns the pressure $P_\alpha$ and 
hydrodynamic 3-velocity ${\bf v}_\alpha$, i.e. those quantities that 
are required for the hydrodynamic transport and are non-additive: 
\begin{eqnarray}
\label{P-uni} 
\widetilde{P}_\alpha= \vartheta P_\alpha + 
(1-\vartheta)\frac{n_\alpha}{n_{\scr{tot}}}P_{\scr{tot}}, 
\\
\label{v-uni} 
\widetilde{{\bf v}}_\alpha= \vartheta {\bf v}_\alpha + 
(1-\vartheta){\bf v}_{\scr{tot}}. 
\end{eqnarray}
Here $\widetilde{P}_\alpha$ and $\widetilde{{\bf v}}_\alpha$ are 
interpolated quantities for the $\alpha$ fluid, and $P_{\scr{tot}}$, 
${\bf v}_{\scr{tot}}$, $n_{\scr{tot}}$, etc. are quantities derived from the 
total baryon density 
$J_{\scr{tot}}^{0}=J_{\scr p}^{0}+J_{\scr t}^{0}$ and the $0\nu$
  components of the total baryon-rich energy--momentum tensor 
$T^{0\nu}_{\scr{tot}}=T^{0\nu}_{\scr p}+T^{0\nu}_{\scr t}$, 
assuming that p- and t-fluids are unified. 
The difference between $P_\alpha$ and $P_{\scr{tot}}$, and
between ${\bf v}_\alpha$ and ${\bf v}_{\scr{tot}}$, respectively, is
as follows. From the solution of hydrodynamic equations
(\ref{eq8})--(\ref{eq8t}) and (\ref{eq13}) at each time step we obtain
hydrodynamic quantities $J_{\alpha}^{0}$, $T^{0\nu}_\alpha$ and 
$T^{\scr{(eq)}0\nu}_{\scr f}$. At the same time, we need,  
in particular, the
pressure and 
hydrodynamic 4-velocity in order to proceed to the next time step. We
have to calculate them based on the above  hydrodynamic
quantities and the EoS. Here we can proceed in two ways. 
The conventional way of the multi-fluid hydrodynamics consists in
calculating the pressure and 4-velocity for each fluid, i.e. solely
based on the $J_{\alpha}^{0}$ and $T^{0\nu}_\alpha$ quantities
related to this fluid. This way we arrive at $P_\alpha$ and 
${\bf v}_\alpha$, which completely preserve the multi-fluid character
of the solution. Alternatively, we can completely abandon the
multi-fluid nature and assume that we deal with a locally equilibrium
piece of baryon-rich matter, which is characterized by
$J_{\scr{tot}}^{0}$ and $T^{0\nu}_{\scr{tot}}$, and then calculate the
pressure and 4-velocity of this ``unified'' piece. Thereby we arrive
at $P_{\scr{tot}}$ and ${\bf v}_{\scr{tot}}$, which describe unified
baryon-rich fluids as if they are mutually stopped.

This unification procedure was first proposed by the 
Los-Alamos group \cite{Amsden} and then used in subsequent applications 
of the multi-fluid dynamics \cite{MRS91,INNTS,Kat93,Brac97}. 
Evidently, this procedure also affects the stopping power and 
therefore can be added to 
the above list of uncertainties associated with the friction
force $F_{\alpha}^\nu$. 
Other criteria of unification may
produce different observable stopping, as it was shown
in Ref. \cite{Brac00a}. However, we prefer to keep the original
criterion, keeping in mind that this is an important part of the
stopping prescription. 

As compared to the already conventional unification procedure
described above, we have also introduced $\vartheta^2$ factor in
the friction force itself, cf. Eq. (\ref{eq16}). 
It is justified, since only the
$\vartheta$ fraction of each baryon-rich fluid remains in the two-fluid
regime, as it follows from the above discussion. Therefore, in order to 
be consistent with the above unification procedure, we should keep only 
this $\vartheta$ fraction of the density of each fluid, 
$\rho^{\xi}_{\alpha}$, in the friction term. 

As it has been already mentioned, the way, in which the unification 
procedure is realized, affects the stopping of the nuclear 
matter. In fact, it is possible to avoid this artificial unification 
procedure. As it was shown in Refs. \cite{gsi94,I87}, the formulation 
of the multi-fluid dynamics in terms of mean fields rather than of 
the EoS results in automatic unification of mutually stopped matter. 
However, the problem of such mean-field formulation is that the mean 
fields become unrealistically strong at high relative velocities of 
counter-streaming matter, whereas they should die out due to momentum 
dependence of self energies, as it was  advocated in the HSD model Ref.
 \cite{Cassing99}. Therefore, for the domain of high incident energies 
the present approach, based on a separate EoS in each fluid and  
complemented by the unification procedure, is certainly preferable 
as compared to the pure mean-field formulation of Refs. 
\cite{gsi94,I87}. While the latter formulation is definitely 
advantageous at moderate incident energies of the order of 1$A$ GeV.

Eqs.~(\ref{eq8})--(\ref{eq8t}) and (\ref{eq12}), supplemented by a
certain EoS and expressions for friction forces $F^\nu$, form a
full set of equations of the relativistic 3-fluid hydrodynamic
model. The only quantity, we still need to define in terms of
hydrodynamic variables and some cross sections, is the friction of the
fireball fluid with the p- and t-fluids, 
$F_{\scr{fp}}^\nu$ and $F_{\scr{ft}}^\nu$.

\subsection{Interaction between Fireball and Baryon-Rich Fluids}
\label{Interaction between Fireball and Baryon-Rich Fluids}

Our aim here is to estimate the scale of the friction force
between the fireball and baryon-rich fluids, similar to that done
before for baryon-rich fluids \cite{Sat90}. To this end, we
consider a simplified system, where all baryon-rich fluids consist
only of nucleons, as the most  abundant component of these
fluids, and the fireball fluid contains only  pions.

For incident energies from 10$A$ (AGS) to 200$A$ GeV (SPS), the
relative nucleon-pion energies are in the resonance range
dominated by the $\Delta$-resonance. To estimate this relative
energy we consider a produced pion, being at rest in the center of
mass of the colliding nuclei, $p=\{m_\pi,0,0,0\}_{cm}$.
Baryon-rich fluids decelerate each other during their
inter-penetration. This means that the nucleon momentum $q$ should be
smaller than the incident momentum, $q_0 = E_N <m_N\gamma_{cm}$, 
where $\gamma_{cm}$ is the
gamma factor of the incident nucleon in the c.m. frame. Calculating
the invariant relative energy squared $s=(p+q)^2$ at
$E_{\scr{lab}}=$ 158$A$ GeV, we obtain $s^{1/2}<$ 1.8 GeV.
This range of $s$ precisely covers the resonance region, 1.1 GeV
$<s^{1/2}<$ 1.8 GeV \cite{PPVW93}. At $E_{\scr{lab}}=$ 10$A$ 
GeV we arrive at $s^{1/2}<$ 1.3 GeV, which is also within
the resonance region. At even lower incident energies the strength
of the fireball fluid becomes so insignificant, as compared with
thermal mesons in the p- and t-fluids, that the way of treatment
of its interaction with the baryon-rich fluids does not
essentially affect  the observables. For the same reason we do not
apply any special prescription for the unification of the fireball
fluid with the baryon-rich fluids, since this may happen only at
relatively low incident energies $E_{\scr{lab}} <$ 10$A$ GeV.

The resonance-dominated interaction implies that the essential
process is absorption of a fireball pion by a p- or t-fluid
nucleon with formation of an $R$-resonance (most probably
$\Delta$). This produced $R$-resonance still belongs to the
original p- or t-fluid, since its recoil due to absorption of a
light pion is small. Subsequently this $R$-resonance decays into a
nucleon and a pion already belonging to the original p- or
t-fluid. Symbolically, this mechanism can be expressed as $$
N^\alpha + \pi^{\scr f} \to R^\alpha \to  N^\alpha + \pi^\alpha
.$$ As a consequence,  only the loss term contributes to the
kinetic equation for the fireball fluid.

Proceeding from the above consideration, we write down the
collision term between fireball-fluid pions and  $\alpha$-fluid
nucleons ($\alpha=$p or t) as follows
\begin{eqnarray}
\label{eq18} C_{\scr f} (f_\alpha,f_{\scr f}) = - \!\int \frac{d^3
q}{q_0} \ W^{N\pi\to R}(s) \ f_{\scr f}^{\scr{(eq)}} (p)
f_\alpha (q),
\end{eqnarray}
where $s=(p+q)^2$, $$W^{N\pi\to R}(s) = \frac{1}{2}
\sqrt{(s-m_N^2-m_\pi^2)^2-4m_N^2m_\pi^2}
\sigma_{\scr{tot}}^{N\pi\to X}(s)$$ is the rate to produce a
baryon $R$-resonance, and $\sigma_{\scr{tot}}^{N\pi\to X}(s)$ is
the parameterization of experimental pion--nucleon cross-sections
\cite{PPVW93}. Here, only the distribution function of formed (and
hence thermalized) fireball pions, $f_{\scr f}^{\scr{(eq)}}$,
enters the collision term, since the non-formed particles do not
participate in the interaction by assumption.

Multiplying $C_{\scr f} (f_\alpha,f_{\scr f})$ by the 4-momentum $p^\nu$
and integrating the result over momentum, we arrive at
\begin{eqnarray}
&&F_{\scr{f}\alpha}^\nu (x) = \int \frac{d^3
q}{q_0} \frac{d^3 p}{p_0} p^\nu  W^{N\pi\to R}(s) \ f_{\scr f}^{\scr{(eq)}}
(p)  \ f_\alpha (q) 
\cr 
&&\simeq \frac{W^{N\pi\to
R}(s_{{\scr f}\alpha})}{m_\pi u_{\scr f}^0} \left(\int \frac{d^3
q}{q_0} f_\alpha (q)\right) \left(\int \frac{d^3 p}{p_0} p^0 p^\nu
f_{\scr f}^{\scr{(eq)}} (p)\right)
\cr
&&=
 D_{\scr{f}\alpha}\frac{T^{\scr{(eq)}0\nu}_{\scr f}}{u_{\scr
 f}^0}\rho_{\alpha},
\label{eq19}
\end{eqnarray}
where we substituted $p^0$ and $s$ by their mean values,
$\langle p^0\rangle =m_\pi u_{\scr f}^0$ and $s_{{\scr f}\alpha} = (m_\pi
u_{\scr f}+m_N u_{\alpha})^2$, and introduced the transport
coefficient
\begin{eqnarray}
D_{\scr{f}\alpha} = \frac{W^{N\pi\to R}(s_{{\scr f}\alpha})}{m_N m_\pi}
=V_{\scr{rel}}^{{\scr f}\alpha} \ \sigma_{\scr{tot}}^{N\pi\to
R}(s_{{\scr f}\alpha}).
\end{eqnarray}
Here, $V_{\scr{rel}}^{{\scr f}\alpha}=[(s_{{\scr
f}\alpha}-m_N^2-m_\pi^2)^2 -4m_N^2m_\pi^2]^{1/2}/(2m_N m_\pi)$
denotes the mean invariant relative velocity between the fireball
and the $\alpha$-fluids. Thus, we have expressed the friction
$F_{\scr{f}\alpha}^\nu$ in terms of the fireball-fluid
energy-momentum density $T^{0\nu}_{\scr f}$ 
(of only pions as yet), 
the scalar density
$\rho_{\alpha}$ of the $\alpha$-fluid (of only nucleons as yet), 
and a transport coefficient
$D_{\scr{f}\alpha}$. Note that this friction is zero until the
fireball pions are formed, since $T^{\scr{(eq)}0\nu}_{\scr f}=0$
during the formation time $\tau$.

 In fact, the above treatment is
an estimate of the friction terms rather than their strict
derivation. All the uncertainties mentioned in the previous
subsection are well applied to the case under consideration. This
peculiar way of evaluation is motivated by the form of the final
result (\ref{eq19}). An advantage of this form is that $m_\pi$ and
any other mass do not appear explicitly, and hence it allows a
natural extension  to any content of the fluid, including
deconfined quarks and gluons, assuming that $D_{\scr{f}\alpha}$
represents just a scale of the transport coefficient. 
Performing such extension, we assume that $T^{0\nu}_{\scr f}$
represents the total energy-momentum density of the
fireball-fluid and $\rho_{\alpha}$ is the total scalar density
of the $\alpha$-fluid, i.e. $\rho_{\alpha}\equiv\rho_{\alpha}^{\xi=1}$ 
in terms of Eq. (\ref{ro_xi}). 
Here we have omitted tuning $\xi$ factors in the scalar density
$\rho_{\alpha}$, since in this case their effect is very similar
to that of the formation time, which switches on/off the
interaction at various stages and thereby effectively changes its
strength.

\subsection{Freeze-Out}
\label{Freeze-Out}

The hydrodynamic simulation is terminated by a freeze-out
procedure. Though this method (as applied to high-energy physics)
was first proposed almost 50 years ago \cite{Milekhin}, this is still
an actual 
problem which is actively discussed. The method  was intuitively
clear and easily applicable. However, as was shown by Cooper and
Frye \cite{Cooper}, the Milekhin's method violates the energy
conservation. To remedy the situation, they proposed their own
recipe. The Cooper--Frye recipe \cite{Cooper} was not free of
problems as well. It gives negative contribution to the particle
spectrum in some kinematic regions in which the normal vector to
the freeze-out hyper-surface is space-like. This negative
contribution corresponds to frozen out particles returning to the
hydro phase. Cut off of this negative contribution again returns us to
the violation of the energy conservation. 
The get rid of this negative spectrum, there was
proposed a modification of the Cooper--Frye recipe based on a
cut-J\"uttner distribution
\cite{Bugaev96,Neumann97,Csernai97,Bugaev99}. In this distribution
the part of the J\"uttner distribution that gave the negative
spectrum is simply cut off. To preserve the particle and energy
conservation, the rest of J\"uttner distribution is renormalized,
effectively resulting in a new temperature and chemical potential
(so called "freeze-out shock"). In fact, this cut-J\"uttner recipe
has no physical justification, except for practical utility.
Moreover, the cut-J\"uttner recipe is not supported by schematic
kinetic treatment \cite{Csernai99} of the transitional region from
hydro regime to that of dilute gas and  looks like a violence to
the nature, making the freeze-out procedure uncontrollable.
Recently there was proposed a new freeze-out recipe, a
cancelling-J\"uttner distribution \cite{Csernai04}, which complies
with results of schematic kinetic treatment \cite{Csernai99}. 
However, very recently the authors reported \cite{Csernai05} that this 
cancelling-J\"uttner distribution is satisfactory only for the
space-like freeze-out, while it fails for the time-like one. 
It should be stressed that this was precisely the schematic
kinetic treatment. This region, where the transition from highly
collisional dynamics to the collisionless one occurs, is highly
difficult for the kinetic treatment and hardly allows any
justified simplifications.

All the above considerations of the freeze-out process proceeded
from assumption of existing some continuous hyper-surface separating
the hydro system from the frozen-out gas. Conservation conditions
on such hyper-surface are constructed in analogy with shock front in
hydrodynamics. From the practical point of view, it would mean that
we should first run the hydro calculation without any freeze-out
and only after that look for a hyper-surface, where the freeze-out
criterion is met. In practice our hydro simulation proceeds in
different way. The freeze-out criterion is checked continuously
during the simulation. If some parts of the hydro system meet this
criterion, they decouple from the hydro calculation. 
The frozen-out matter escapes from the system, removing all the energy
and momentum accumulated in this matter. Therefore, it produces no
recoil to the rest of still hydrodynamic system%
\footnote{
In fact, the particle emission would create pressure against the 
emitting surface, if we allow  feedback between the frozen-out matter 
and the still hydrodynamic system, i.e. if some particles from the 
frozen-out matter return into the hydrodynamic system. This is a 
long-standing problem of the freeze-out procedure still waiting for 
its consistent solution. 
}. 
In particular, it means that the boundary condition on the free
surface (between the hydro system and vacuum) is not kept at the
same position as in the calculation without freeze-out but moves
inside the system. It affects not only the system surface but also
the interior. The freeze-out process looks like an evaporation (or
fragmentation, on account of final-size grid) first from the
system surface and then as a volume fragmentation of the system
residue.

In view of above said, we prefer to avoid not quite justified
complications of the freeze-out procedure and make use of the
simplest (however equally unjustified) original choice of Milekhin
\cite{Milekhin} with corrected treatment of the energy
conservation. The freeze-out criterion we use is
\begin{eqnarray}
\label{FOcriterion} \varepsilon_{\scr{tot}} <
\varepsilon_{\scr{frz}},
\end{eqnarray}
where
\begin{eqnarray}
\label{eps_tot} \varepsilon_{\scr{tot}} = \left(T^{00}_{\scr p} +
T^{00}_{\scr t} + T^{\scr{(eq)}00}_{\scr f}\right)_{\scr{proper}}
\end{eqnarray}
is the energy density of all three fluids in the proper reference
frame, where all nondiagonal components of the total
energy--momentum tensor are zero, 
\begin{eqnarray}
\label{proper_tot} \left(T^{\mu\nu}_{\scr p} + T^{\mu\nu}_{\scr t}
+ T^{\scr{(eq)}\mu\nu}_{\scr f}\right)_{\scr{proper}}^{\mu\ne\nu}=0,
\end{eqnarray}
and $\varepsilon_{\scr{frz}}$ is the critical freeze-out energy
density. 
If the freeze-out criterion is met in some space--time point, all
three fluids in this point get frozen out and removed from the
hydrodynamic evolution. In fact, we freeze out the fluids in tiny
portions, i.e. droplets. This is allowed and even implied by the
numerical scheme we use (see App. \ref{Numerics}). Each
droplet gets frozen out in its proper reference frame. 
In terms of
the hypersurface, our freeze-out hypersurface is discontinuous and
consists of small fragments with normal vectors of local
4-velocities. 
Any transport is prohibited on the non-existent parts 
(parallel to the flow velocity) of the surface. 
In this sense, our freeze-out is more similar to a
continuous fragmentation of the system rather than to
occurrence of the shock front.

A frozen-out droplet of the $\alpha$-fluid (let it be marked as
$i\alpha$) is still characterized by some temperature, baryon and
strange chemical potentials corresponding to a nongas EoS
(involving some mean fields) used in the hydro calculation. This
is not suitable for calculation of the spectrum of observable
particles. First we should release the energy stored in mean
fields. To do this, we recalculate temperature
($T^{i\alpha\scr{(gas)}}$), baryon ($\mu_b^{i\alpha\scr{(gas)}}$)
and strange ($\mu_s^{i\alpha\scr{(gas)}}$) chemical potentials
corresponding to the hadronic gas EoS proceeding from
conservations of total energy--momentum, baryon and strange charges
in the droplet. 
This has been done in the following way. From the solution of 
hydrodynamic equations (\ref{eq8})--(\ref{eq8t}) and (\ref{eq13})
we know five quantities for each fluid: $J^0_\alpha$, 
$T^{00}_\alpha$, $T^{01}_\alpha$, $T^{02}_\alpha$ and $T^{03}_\alpha$. 
In order to interpret them in thermodynamic terms, we should first
determine six 
quantities for each fluid: $n_\alpha$, $\varepsilon_\alpha$, $P_\alpha$ and 
3 components of the 4-velocity $u_{\alpha}$, 
proceeding from above five hydrodynamic quantities, see Eqs.  
(\ref{eq11}) and (\ref{eq12}). Naturally, five equations are not enough 
for determining six quantities. Therefore, we add one more equation to this
set---the EoS. 
Thus, the resulting thermodynamic quantities turn out to be 
EoS dependent. In particular, this scheme with the gas EoS 
results in a change of the hydrodynamic 4-velocity
($u^\mu_{i\alpha\scr{(gas)}}$) as compared to that calculated 
with the EoS used in the hydrodynamic simulation. 
The second, already conventional step consists in determination 
of temperature and chemical potentials proceeding  
from baryonic, strange and energy densities and pressure.

This is a kind of
``freeze-out shock'', which however is completely different from
that induced by the cut-J\"uttner recipe. Now in terms of
frozen-out droplets of various $\alpha$-fluids,
  the spectrum of observable hadrons of $a$ species with
$e^a_b$ baryon and $e^a_s$ strange charges can be expressed as
follows
\begin{widetext}
\begin{eqnarray}
\label{FOspectrum} E \frac{d N_a}{d^3 p} = \sum_{i\alpha}
\frac{g_a V_{i\alpha}^{\scr{(proper)}}}{(2\pi)^3}
\frac{p_\mu u^\mu_{i\alpha\scr{(gas)}}}%
{\exp\left\{\left(p_\mu u^\mu_{i\alpha\scr{(gas)}} - e^a_b
\mu_b^{i\alpha\scr{(gas)}} - e^a_s
\mu_s^{i\alpha\scr{(gas)}}\right)/T^{i\alpha\scr{(gas)}}\right\}
\pm 1}
\end{eqnarray}
\end{widetext}
where $g_a$ is degeneracy of the $a$ particle,
$V_{i\alpha}^{\scr{(proper)}}$ is the volume of the 
$i\alpha$-droplet in its rest frame, and the sum runs over all
frozen-out droplets of all fluids. If the $a$ species is a baryon,
the upper sign ($+$) should be taken in the denominator, if it is
a meson, the lower sign ($-$).

The described ``fragmentation'' method of freeze-out precisely
conserves energy--momentum and various charges. However, it is
also not free from problems. The microscopic justification of this
method is still lacking. The problem of returning frozen-out
particles into the hydrodynamic phase still persists. Certainly,
further search for a reliable freeze-out procedure is needed. A more
consistent way of performing the freeze-out, the method of ``continuous
emission'', was proposed in Ref. \cite{Sinyukov02}. 
Avoiding sharply defined freeze-out hypersurface, this method considers
a continuous 
emission of particles from a finite volume, governed by their mean
free paths. Its predictions at the level of observables differ from
those based on the Cooper--Frye recipe \cite{Grassi04}. 
Unfortunately, this method is very difficult for the numerical
implementation.

\section{Simulations of Nucleus--Nucleus Collisions}

The strategy of our simulations is as follows. 
At the
first step, we try to reproduce the stopping power observed in
proton rapidity distributions by means of fine fitting of friction
forces of the model. In principle, friction forces are EoS
dependent both through scalar densities (\ref{ro_xi}) and 
because of medium modifications of cross sections. In
spite of this, our friction is 
just schematically estimated proceeding from vacuum proton--proton
cross sections and therefore does not comply with the EoS used.
This is so even for the simplest hadronic EoS, used in the present
paper. The list of uncertainties relevant to the friction forces has
been discussed in Subsect.  \ref{Friction between Baryon-Rich Fluid}. 
In view of this, the strategy of
fine fitting of friction forces is quite reasonable. Note that the
success of such fit is not obvious in advance because by means of
two functions $\xi_h (s)$ and $\xi_q(s)$ of a single variable (in
the present case, only $\xi_h(s)$), see Eq. (\ref{ro_xi}), we fit a
function of three variables: rapidity, incident energy and impact
parameter. It is worthwhile to mention that the fit of the
friction is EoS dependent. It allows to reproduce proton rapidity
distributions only with the particular EoS. In general, another EoS
requires different fit.

After the friction forces have been fixed, we tune the freeze-out
energy density (in the reasonable 
range) in order to reproduce transverse-mass proton spectra and 
multiplicities of produced pions at comparatively low incident
energies $E_{\scr{lab}}\lsim$ 20$A$ GeV. At higher incident
energies, the pion multiplicity 
turns out to be weakly sensitive to
variation of the freeze-out energy density. 
In principle, the freeze-out criterion could be different for different
particle species and even for the chemical and thermal
freeze-out. However, we keep it unique for all the cases 
in order to avoid multiplication of fitting parameters. Therefore, fit of
these quantities by means of a single parameter is not an obvious
task. 
At higher incident energies $E_{\scr{lab}}\gsim$ 30$A$ GeV
the pion multiplicities become
sensitive to the formation time $\tau$. Therefore, the next step
consists in tuning $\tau$ (again in the reasonable range) to reproduce
them already at higher incident energies.
These subsequent steps
are simplified by the fact that proton rapidity distributions are
only slightly sensitive to variation of $\tau$ and the freeze-out
energy density. 
After all these steps, all the model parameters
got fixed, and all further calculations give pure predictions of
the model. Our final aim is to find a EoS which reproduces in the best
way 
the largest body of the observables
of nuclear collisions in the
incident energy range $E_{\scr{lab}}\simeq$ (1--160)$A$ GeV.

\subsection{Hadronic EoS}
\label{Hadronic EoS}

We start our simulations with the purely hadronic EoS. This
EoS, which was originally used in 2-fluid simulations
\cite{MRS88,MRS91,INNTS}, was proposed in \cite{gasEOS}. It is a
natural reference point for any other more elaborate EoS. The
energy density and pressure are constructed as follows:
\begin{eqnarray}
\label{E} 
\hspace*{-5mm}
\varepsilon (n_B,T)&=&
\varepsilon_{\scr{gas}}(n_B,T)+W(n_B),
\\
\label{P}
\hspace*{-5mm}
P(n_B,T)&=&P_{\scr{gas}}(n_B,T)+n_B\frac{dW(n_B)}{dn_B}-W(n_B),
\end{eqnarray}
where $\varepsilon_{\scr{gas}}(n_B,T)$ and $P_{\scr{gas}}(n_B,T)$
are the energy density and pressure of relativistic hadronic gas,
respectively, which depend on baryon density $n_B$ and temperature
$T$. The only difference from the ideal gas is that baryons are
affected by a mean field $U(n_B)$, i.e. the energy of the $a$-baryon
of mass $M_a$ with momentum ${\bf p}$ is  
$\epsilon_a=({\bf p}^2+M_a^2)^{1/2}+b_a U(n_B)$, where $b_a$ is the
baryon number of the $a$-particle, and the potential
$U(n_B)$ is parametrized as follows
\begin{eqnarray}
\label{U(n)} 
U(n_B)= m_{N}\left[-2b\left(\frac{n_B}{n_{0}}\right)
+ c(\gamma +2)\left( \frac{n_B}{n_{0}}\right) ^{\gamma +1}\right]. 
\end{eqnarray}
It depends only on the density $n_B$. The self-consistent potential
contribution to the energy density, $W(n_B)$, is 
\begin{eqnarray}
\label{W} 
W(n_B)= \int_0^{n_B} U(n) \ dn. 
\end{eqnarray}
Parameters $b$, $c$ and $\gamma$ are determined from the
condition that the cold nuclear matter saturates at $n_0=$ 0.15
fm$^{-3}$ and $\varepsilon (n_0,T=0)/n_0 - m_N=-16$ MeV, and
incompressibility of this nuclear matter is $K=$ 210 MeV. 
\begin{figure}[ht]
\includegraphics[width=7cm]{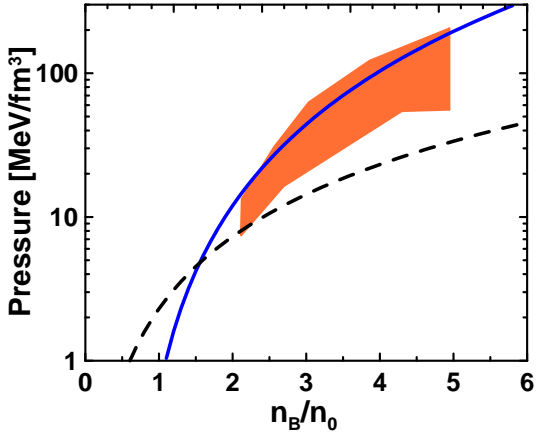}
$\;$\vspace*{-2mm} 
\caption{ (Color online) Baryon-density dependence of the pressure at
  $T=0$. Shaded region is the constraint \cite{DLL02} derived from
  experimental data. Solid and dashed lines present the 
  pressure $P(n_B,T=0)$ and that for the ideal hadronic gas, 
  $P_{\scr{gas}}(n_B,T=0)$ (cf. Eq. (\ref{P})), respectively.} 
\label{fig0}
\end{figure}
Parametrization (\ref{U(n)}) results in superluminal sound velocity
at densities $n_B/n_0>$8. To preserve causality at high $n_B$, the
following form of the energy density
\begin{eqnarray}
\varepsilon (n_B,T\!=\!0)= n_{0}m_{N}\left[ A\left(
\frac{n_B}{n_{0}}\right)^{2}+C+B\left( \frac{n_{0}}{n_B}\right)
\right]
\end{eqnarray}
is used at $n_B/n_0 >$ 6. Parameters $A$, $B$ and $C$ are
determined on the condition that $\varepsilon (n_B,T=0)$ and its
two first derivatives are continuous at $n_B/n_0 =$ 6.
As seen from Fig. \ref{fig0}, the pressure of this hadronic EoS 
is within the constraint given by Danielewicz et
al. extracted from 
the analysis of flow of  nuclear matter \cite{DLL02} at the AGS
incident energies.

\subsection{Summary of Parameters: Friction, Freeze-Out, etc.}
\label{Summary of Parameters}

For the sake of complete account, 
in this section we would like to summarize the parameters used in
the present simulations. The specific reasons for choosing these parameters
will be discussed in subsequent Subsections. 

\begin{itemize}
\item
The key quantity is the EoS, which was taken in the simple form
of purely hadronic EoS, see Sect. \ref{Hadronic EoS}.
In this EoS, 48 different hadronic species are taken into account. 
Each hadronic species includes all the relevant isospin states, e.g., the
nucleon species includes proton and neutron. 
\item
The friction between baryon-rich fluids was fitted to reproduce
the stopping power observed in proton rapidity distributions, 
see Sect. \ref{Proton Rapidity}. For
this purpose the original friction, parametrized through experimental
inclusive proton--proton cross sections \cite{Sat90}, was enhanced
by means of the tuning factor $\xi_h$
\begin{eqnarray}
\label{xi} \xi_h^2 (s) =\gamma_h+2\beta_h
\left[\ln\left(s/(2m_N)^2\right)^{1/2}\right]^{1/4}
\end{eqnarray}
with $\gamma_h=$ 1 and $\beta_h=$ 0.75, see Fig. \ref{fig15}. 
The $\xi_q$ factor is not applicable 
here because of the pure hadronic nature of the EoS.
\begin{figure}[ht]
\includegraphics[width=7cm]{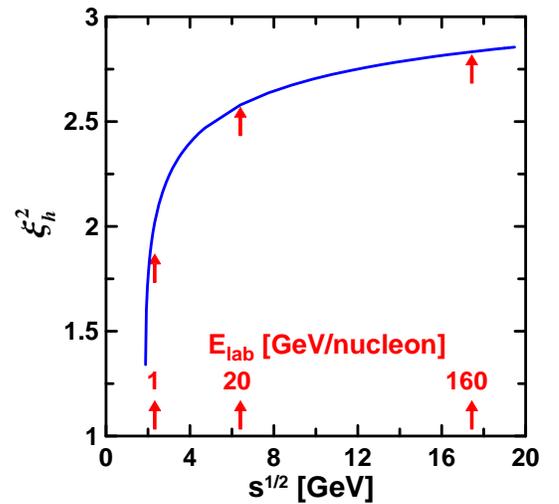}
\caption{(Color online) Fitted friction enhancement for the purely
  hadronic EoS \cite{gasEOS} used in simulations, see Eq. (\ref{ro_xi}).}
\label{fig15}
\end{figure}
\item
The parameter of freeze-out energy density was fitted to reproduce 
transverse-mass proton spectra, see Subsect. \ref{Proton Transverse}, and 
multiplicities of produced pions at comparatively low incident
energies $E_{\scr{lab}}\lsim$ 20$A$ GeV, 
see Subsect. \ref{Pion Rapidity}. It has been taken
$\varepsilon_{\scr{frz}}\simeq$ 0.2 GeV/fm$^3$ at all incident energies
with the only exception: $\varepsilon_{\scr{frz}}\simeq$ 0.1 GeV/fm$^3$
at $E_{\scr{lab}} \leq$ 2$A$ GeV. These $\varepsilon_{\scr{frz}}$
values are presented as approximate quantities because of the numerical
realization of the freeze-out procedure (cf. App. A). 
\item
The formation time of the fireball fluid was also fitted to reproduce 
the pion yield at higher incident energies $E_{\scr{lab}}\gsim$ 30$A$ 
GeV, see 
Subsect. \ref{Pion Rapidity}. It has been taken $\tau=$ 2
fm/c. Note that $\tau$ is the formation time of a fluid element,
consisting of 
a number of  particles, which are generally not thermalized. 
This implies that the formation time of a separate particle is
certainly shorter than $\tau$.
\item
The contributions of strong decays of hadronic resonances ($R$) into spectra
of stable hadrons is taken into account, as it is described in App. 
\ref{Resonance Decays}. In the present calculation all spectral
functions were taken without width 
\begin{eqnarray}
\label{AR(s)}
A_R(s) = \delta(s-m_R^2),  
\end{eqnarray}
where $m_R$ is the mass of the $R$ resonance. 
\item
For the reproduction of stable hadron multiplicities it is also important 
to take into account 
contributions of weak decays with small values of $c\tau_w$, where $\tau_w$ is 
the inverse width of the decay. 
Feed back from weak decays of hadrons with $c\tau_w<$ few cm, i.e.  
$K^0_{\scr{short}}$, $\Lambda$, $\Sigma$, $\bar{\Lambda}$ and
$\bar{\Sigma}$\footnote{Note that $\Sigma^0$ and
$\bar{\Sigma}^0$ suffer the sequence of electromagnetic and weak
decays:  $\Sigma^0\to\Lambda\gamma\to N\pi\gamma$.}
have been taken into account in yields of stable particles. 
In particular,  
this is important for reproduction of the pion multiplicities 
(Subsects \ref{Pion Rapidity} and \ref{Multiplicities}) 
and the proper normalization of  
proton spectra (Subsects \ref{Proton Rapidity} and \ref{Proton Transverse}). 
On the other hand, weakly 
decaying hadrons with $c\tau_w$ of the order of few meters, i.e.  
$K^0_{\scr{long}}$, $K^+$ and $K^-$, were treated as stable
particles. 
\item
Light fragment formation (deutrons, tritons, $^3$He and $^4$He) is
taken into account in terms of the coalescence model, which is similar
to that in  Appendix E of Ref. \cite{gsi94}.  In
Ref. \cite{gsi94}, the coalescence coefficients were fitted at the
incident energy 0.8$A$ GeV. To accommodate this formulation
to different incident energies, we change only overall scale of
the coalescence (the coefficient $C_{\scr{coal}}$) which either
enhances ($C_{\scr{coal}}>1$) or reduces ($C_{\scr{coal}}<1$) its
strength, keeping ratios of yields of various fragments the same
as in \cite{gsi94}, see Table \ref{tab:2}. At $E_{\scr{lab}} >$ 9$A$ 
GeV we do not 
apply any coalescence to calculations of nucleon observables,
since the respective correction is negligible. 
\begin{table}[h]
\begin{ruledtabular}
  \begin{tabular}{|c|cccccc|}
$E_{\scr{lab}}$, $A$ GeV& 1 & 2 & 4 & 6 & 8 & $>9$ \\ 
$C_{\scr{coal}}$            &1.5&1.2&0.8&0.4&0.1& 0    \\
  \end{tabular}
\caption{Coalescence parameters $C_{\scr{coal}}$
used for simulations of Au+Au and Pb+Pb
collisions at various incident energies $E_{\scr{lab}}$.
}
\label{tab:2}
\end{ruledtabular}
\end{table}
\end{itemize} 
 
In order to compare with available experimental data, 
impact parameters used in calculations were taken either from
experimental works, if they were evaluated there, or estimated based
on the experimentally declared percentage of the total reaction cross
section, corresponding to a particular event selection.

\subsection{Proton Rapidity Distributions: Observable Stopping Power}
\label{Proton Rapidity}

\begin{figure}[b]
\hspace*{-14mm}
\includegraphics[width=8.0cm]{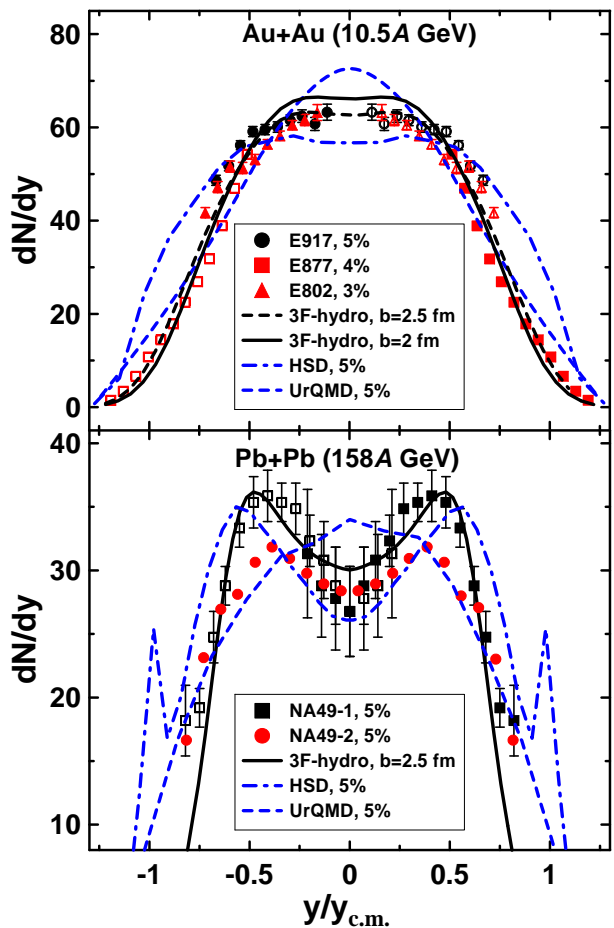}
 \caption{(Color online) Rapidity spectra of protons (upper panel)
and $(p-\bar p)$ (lower panel) from central heavy-ion
collisions. 
Bold solid lines and bold dashed line (upper panel) 
correspond to 3-fluid
hydrodynamic calculations at different impact parameters. 
Thin dashed-dotted and short-dashed lines are the appropriate UrQMD and HSD
model results~\cite{WBCS03}, respectively. Full symbols
display measured experimental points, whereas the open ones are 
those reflected with respect to the mid rapidity point. 
Experimental data are taken by 
collaborations E802~\cite{E802}, E877~\cite{E877},
E917~\cite{E917}, NA49~\cite{NA49-1,NA49-2,NA49-04}. 
The percentage shows the fraction of the total reaction cross section, 
corresponding to experimental selection of central events.} 
\label{fig1}
\end{figure}
\begin{figure*}[uht]
\hspace*{-15mm}
\includegraphics[width=15cm]{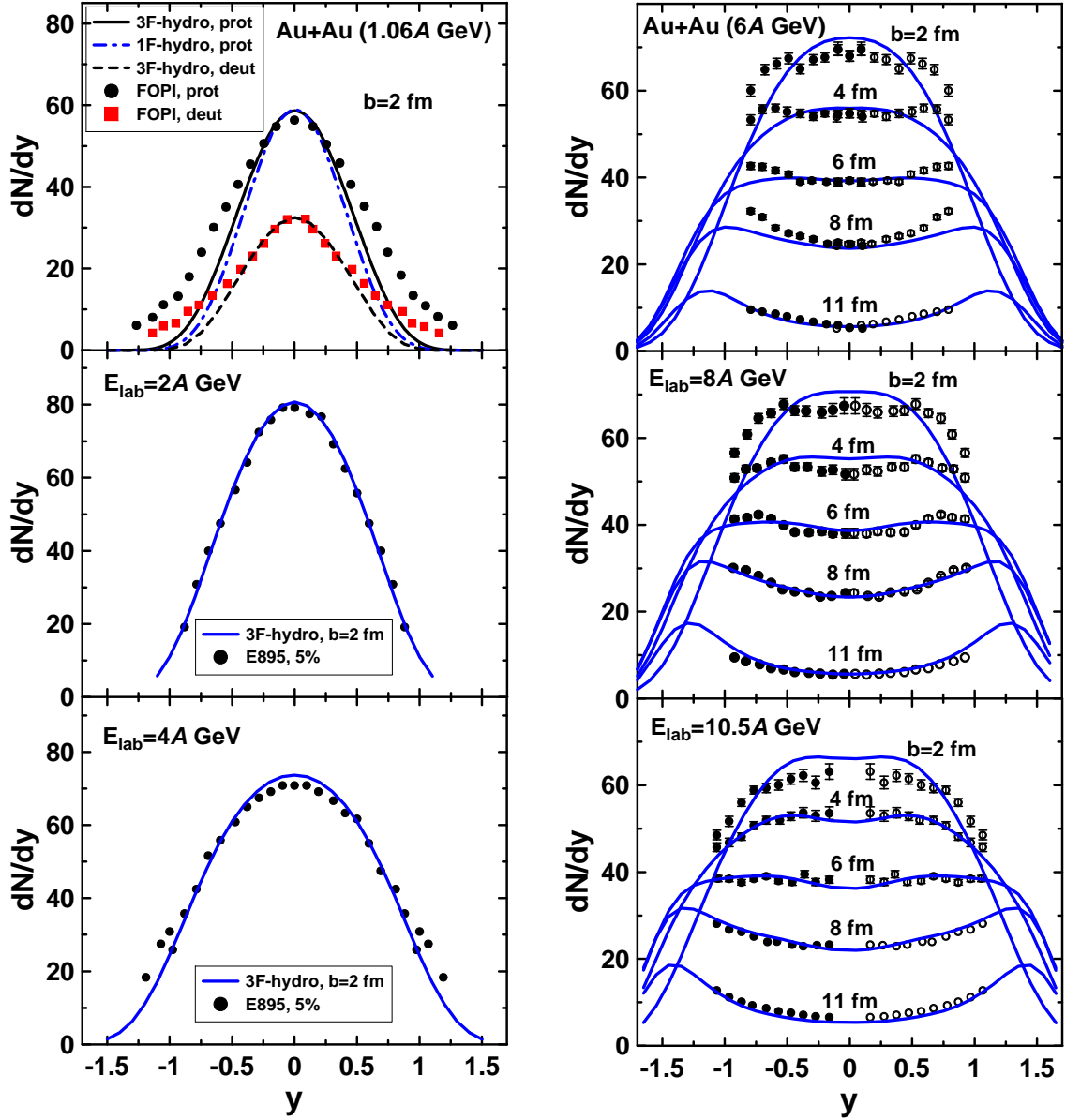}
\caption{(Color online) Proton rapidity spectra at SIS and AGS energies for 
various impact parameter. Solid lines correspond to 3-fluid 
calculations. The dashed line is that for deuteron production. For
comparison the 1-fluid result is shown by the dotted-dashed line at
$E_{\scr{lab}}= 1.06A$ GeV. 
Experimental points at different energies are taken from 
\cite{FOPI} at $E_{\scr{lab}}$ = 1.06$A$ GeV, \cite{E895} at 2$A$ and
4$A$ GeV,  
and \cite{E917} at 6$A$, 8$A$ and 10.5$A$  GeV. The percentage indicates 
the fraction of the total reaction cross section, 
corresponding to experimentally selected events.} 
\label{fig2}
\end{figure*}
\begin{figure}[uht]
\hspace*{-12mm} 
\includegraphics[width=8cm]{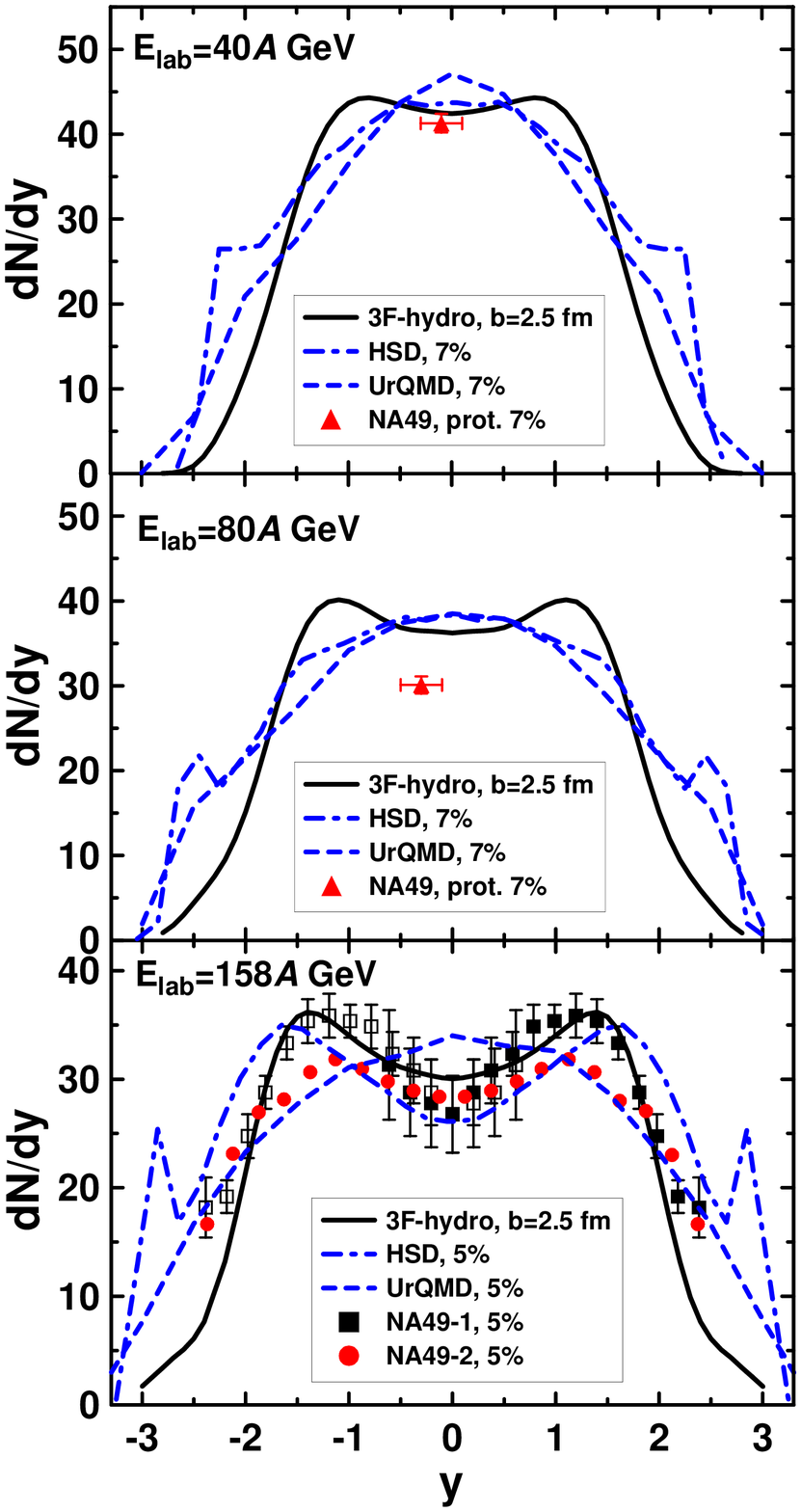}
\caption{ (Color online) $(p-\bar p)$ rapidity distribution
from central Pb+Pb collisions at SPS energies. Solid lines are 3-fluid 
calculations, whereas dashed-dotted
and dashed lines are the corresponding predictions of the 
HSD and UrQMD  models 
\cite{WBCS03}, respectively.  NA49 experimental data are
from Ref.~\cite{NA49-04} for $E_{\scr{lab}}=$ 40$A$ and 80$A$ GeV and
from Refs.~\cite{NA49-1,NA49-2,NA49-04}  for $E_{\scr{lab}}=158A$ 
GeV. The percentage indicates 
the fraction of the total reaction cross section, 
corresponding to experimentally selected events.}
\label{fig3}
\end{figure}

The nucleon rapidity distribution basically reflects the stopping
power achieved in a  nuclear collision. It is defined as   
\begin{eqnarray}
\label{sp_y-mT} 
\frac{dN_N}{dy}=  
\int  d^2 p_T \left[ E \ \frac{dN_N}{ d^3p} + 
\sum_{R} E\frac{d^3 N_N^{(R \to N+X)}}{d^3 p}\right]
\end{eqnarray}
in terms of frozen-out spectra of nucleons $E dN_N/d^3p$, see Eq. 
(\ref{FOspectrum}), and baryonic resonances. 
The second term under the integral takes into account the contribution of 
these resonance decays, cf. Eq. (\ref{R-to-n}), App. \ref{Resonance
  Decays}, the sum runs over all baryonic 
resonances $R$. Integration runs over 
the transverse (with respect to the beam) momentum $p_T$. 

Either identified protons or difference between positive
and negative hadrons are experimentally measured. The latter is associated with
the proton-antiproton difference, $(p-\bar p)$. Since particles are not 
isotopically distinguished in our model, we estimate the proton 
distribution simply as 
\begin{eqnarray}
\label{sp_y-mT-p} 
\frac{dN_p}{dy}=  \frac{Z}{A}\frac{dN_N}{dy}, 
\end{eqnarray}
where $Z$ and $A$ are the proton and mass numbers, respectively, in the 
colliding (identical) nuclei. This estimate is quite reasonable at
comparatively low incident 
energies. At higher energies, when
abundant particle  production starts, this recipe somewhat
underestimates the proton number.  The reason is 
that newly produced particles tend to restore the isotopic symmetry in the 
system and hence increase the number of protons as compared with its
initial value.  The $(p-\bar p)$ quantity
is much less sensitive to the effect of
newly produced particles,  since their contribution is essentially
canceled in the difference $(p-\bar p)$. Therefore, the  $Z/A$ scaling
of the nucleon-antinucleon difference is a reasonable approximation  
for the $(p-\bar p)$ quantity even at high incident energies. 
As mentioned above, our aim is to
reproduce these distributions in a wide range  of incident energies
from 1$A$ to about 160$A$ GeV.

In general, nucleon observables are more robust to variations of
physical parameters than other probes, since they are essentially
confined by the baryon number conservation. 
Our first observation is that the nucleon and, in particular, 
proton rapidity distributions
are only weakly sensitive to variation of the freeze-out energy density
$\varepsilon_{\scr{frz}}$ and the formation time of the fireball fluid $\tau$. 
As for $\varepsilon_{\scr{frz}}$, there is a certain compensation
between effects  of collective motion and internal excitation. Both
these effects produce similar  
consequences in nucleon spectra. For instance, if we allow the system to 
evolve longer (the lower $\varepsilon_{\scr{frz}}$), the collective motion 
of the matter becomes more developed while its internal excitation
drops down (i.e. the matter cools down).  
This counteraction results only in slight changes in 
nucleon rapidity distributions.  
As for the formation time, the density of the produced fireball fluid 
is not high enough even at the highest considered energy. Therefore, its 
interaction with baryon-rich fluids only slightly affects the baryon 
subsystem \cite{3f-yaf}. 
This weak dependence on $\varepsilon_{\scr{frz}}$ and $\tau$ allows us to use 
nucleon rapidity distributions 
to fit the observable stopping power, 
to which nucleon observables are indeed sensitive.

The second main conclusion is that the original friction between
baryon-rich fluids,  
estimated in Ref. \cite{Sat90}
proceeding from free proton--proton cross sections, is 
evidently insufficient to reproduce observable stopping of nuclear matter. 
Therefore, this friction is fitted to reproduce
this stopping power observed in proton rapidity distributions. For
this purpose the original friction was enhanced
by means of the tuning factor $\xi_h$, cf. Eq. (\ref{xi}).
Thus, the friction enhancement is the larger, the higher incident energy is. 
At the lowest considered energy of 1$A$ GeV, the friction turns out to be 
approximately 2 times enhanced. This is similar to earlier results of the  
2-fluid model with mean mesonic fields \cite{gsi94}. There it was
found out that  for the proper reproduction of data on heavy-ion
collisions in the energy range  from 0.4 to 0.8$A$ GeV 
the enhancement factor of 3 was required.

 Fig. \ref{fig1} illustrates the overall quality of reproduction of
 experimental data  by our hydrodynamic calculations as well as by
 other transport simulations  for two basic incident energies of 10
 and 160$A$ GeV.   The 3-fluid simulations were performed at 
 impact parameters which correspond to the experimentally declared
 fraction of the total reaction cross section, accumulated in the
 experimentally selected central  events, assuming sharp cut off in
 impact parameters.  The side bumps of the 
 HSD calculations correspond to spectator parts of colliding
 nuclei. In our model,  
 these spectator parts were cut off, based on simple criterion: the energy 
 per baryon is less than the nucleon mass, which is the case,
 when a nucleon   is at least loosely bound in the matter. 
As seen, experimental data from different experiments somewhat differ,
and therefore one should not expect their reproduction in the 
model calculations within better than $\sim 10\% $. Uncertainties in the
impact parameter at the level of about 0.5 fm do not noticeably affect 
hydrodynamic rapidity spectra. In
general, there is a reasonable agreement between hydrodynamic and
kinetic calculations.

Comparison with experimental data for identified protons is
presented in Fig. \ref{fig2} for the SIS--AGS energy range, where 
various selections of non-central nuclear collisions are also considered. 
Here and below, impact parameters for each set of non-central interactions 
were chosen accordingly the experimentally declared fractions of the
total reaction cross section, accumulated in  the set of
experimentally selected  events, assuming sharp cut off in impact parameters.

At the beam energies $E_{\scr{lab}}$ of the order of 1$A$ GeV, a
large fraction of nucleons is produced with low relative velocities. 
Therefore, they may coalesce forming light fragments. The coalescence
formulation accepted here is similar to that in Appendix E of Ref.
\cite{gsi94}. Only light fragments (deutrons, tritons,
$^3$He and $^4$He) are taken into account. To fit this formulation to
different 
incident energies, we change only overall scale of the coalescence
(the coefficient $C_{\scr{coal}}$) which either enhances
($C_{\scr{coal}}>1$) or reduces ($C_{\scr{coal}}<1$) its strength,
keeping ratios of yields of various fragments the same as in
\cite{gsi94}, see Table \ref{tab:2}.

At the SIS energy we describe reasonably well both proton and
deuteron rapidity spectra. The fact that the experimental distribution 
is somewhat wider than the calculated one can be explained by that 
the set of experimentally selected events in fact contains a certain admixture 
of semi-central and peripheral events, whereas the calculation was performed 
for the single central impact parameter. Indeed, at comparatively low 
energies the reliable selection of central events is problematic \cite{gsi94}. 
This non-central admixture makes the distributions wider as compared to 
what it would be for the perfect central selection. 
At this energy the stopping power is rather
high. This is seen from both the Gaussian-like shape of the rapidity
distributions and also from the fact that 
the 3-fluid results are quite close to those of the conventional 
1-fluid calculations (see Fig. \ref{fig2}). A separate code was used
for these 1-fluid calculations. 
With the increase of the incident energy, 
the spectrum shape starts to 
 differ from the Gaussian one, getting more and more flat at the mid
 rapidity.  The 3-fluid
model reasonably reproduces the dependence of the spectra 
on both the incident energy $E_{\scr{lab}}$ and the impact parameter $b$. 
Note that no extra tuning of normalization of the spectra was done. 
The values of impact
parameters were taken from the experimental estimate of
centrality of nuclear interactions \cite{E917}. Though there are some
uncertainties in this estimate, we applied no special tuning to values
of these impact parameters.

The $(p-\bar p)$ rapidity spectra at the SPS energies are shown
in Fig. \ref{fig3}. A minimum of hydrodynamic distributions $dN/dy$
in the mid rapidity region complies with experimental observations 
at $E_{\scr{lab}}=158A$ GeV. It survives at lower incident energies, 
up to $E_{\scr{lab}}=40A$ GeV. This is in 
contrast to simulations based on kinetic transport codes \cite{WBCS03},
which predict flat or even peacked distributions in the mid rapidity 
region at $E_{\scr{lab}}=$ 40$A$ and 80$A$ GeV. 
Available experimental points here do not allow us to verify 
these different predictions. Note that experimental points at
$E_{\scr{lab}}=$ 40$A$ and 80$A$ GeV correspond to identified protons,
while all presented calculations are related to $(p-\bar p)$.

\subsection{Proton Transverse Mass Distributions}
\label{Proton Transverse}

\begin{figure*}[ht]
\includegraphics[width=13.9cm]{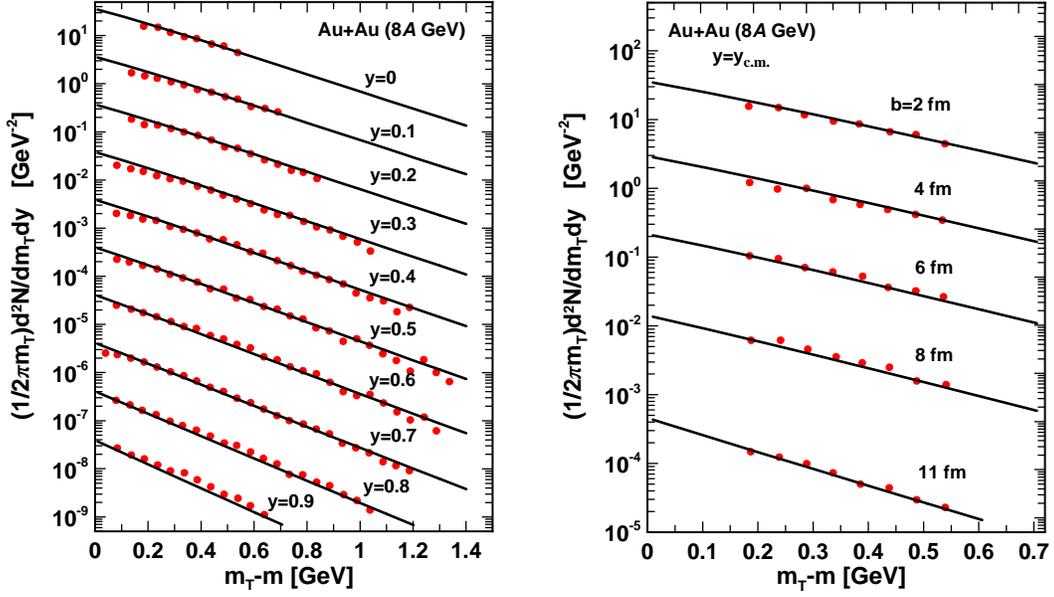}
\caption{(Color online) Transverse mass distributions of protons 
from Au(8$A$ GeV)+Au  collisions at $b=$ 2 fm and different rapidities in
the c.m. system (left panel) and at the mid rapidity and 
different impact parameters (right panel), cf. the right panel of
Fig. \ref{fig2}. For clarity of representation,  
every next data set and the corresponding curve (from top to bottom) is 
multiplied by the additional 
factor 0.1. Experimental data are from \cite{E917}. }
\label{fig4}
\end{figure*}
\begin{figure}[ht]
\includegraphics[width=6.5cm]{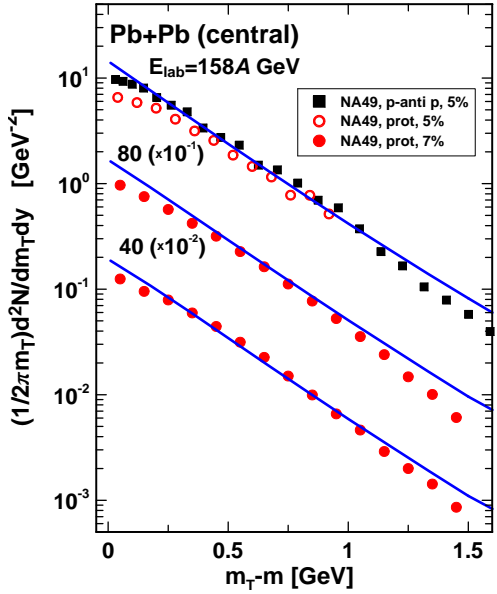}
 \caption{(Color online) Transverse
mass distributions of protons at the mid rapidity from central
Pb+Pb collisions at incident energies 
$E_{\scr{lab}}=$ 158$A$, 80$A$ and 40$A$ GeV and impact parameter $b=$
2.5 fm, cf. Fig. \ref{fig3}. NA49 experimental data are taken
from \cite{NA49-1} (squares), \cite{NA49-03} (open circles) and
~\cite{NA49-03} (full circles). }
\label{fig5}
\end{figure}
\begin{figure}[ht]
\vspace*{-4mm}
\includegraphics[width=8.2cm]{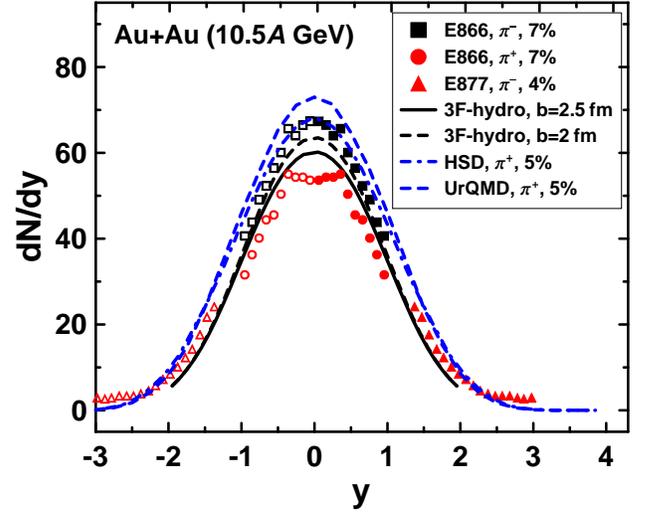}
\vspace*{-3mm}
 \caption{(Color online) Pion rapidity spectra from central Au(10.5$A$
   GeV)+Au  collisions. Bold solid and dashed lines represent 3-fluid
calculations at $b=$ 2.5 and 2.0 fm, respectively.
Thin dashed-dotted and short-dashed lines correspond $\pi^+$ spectra from 
kinetic simulations 
within HSD and UrQMD models \cite{WBCS03}. 
The $\pi^-$ (full squares and triangles) and $\pi^+$ (full circles) data 
are measured by E895~\cite{E895} and E877~\cite{E877}
collaborations. The percentage indicates 
the fraction of the total reaction cross section, 
corresponding to experimentally selected events. Open symbols are
obtained by reflecting the full ones with 
respect to the mid rapidity.} \label{fig8}
\end{figure}

\begin{figure*}[ht]
\includegraphics[height=14.2cm]{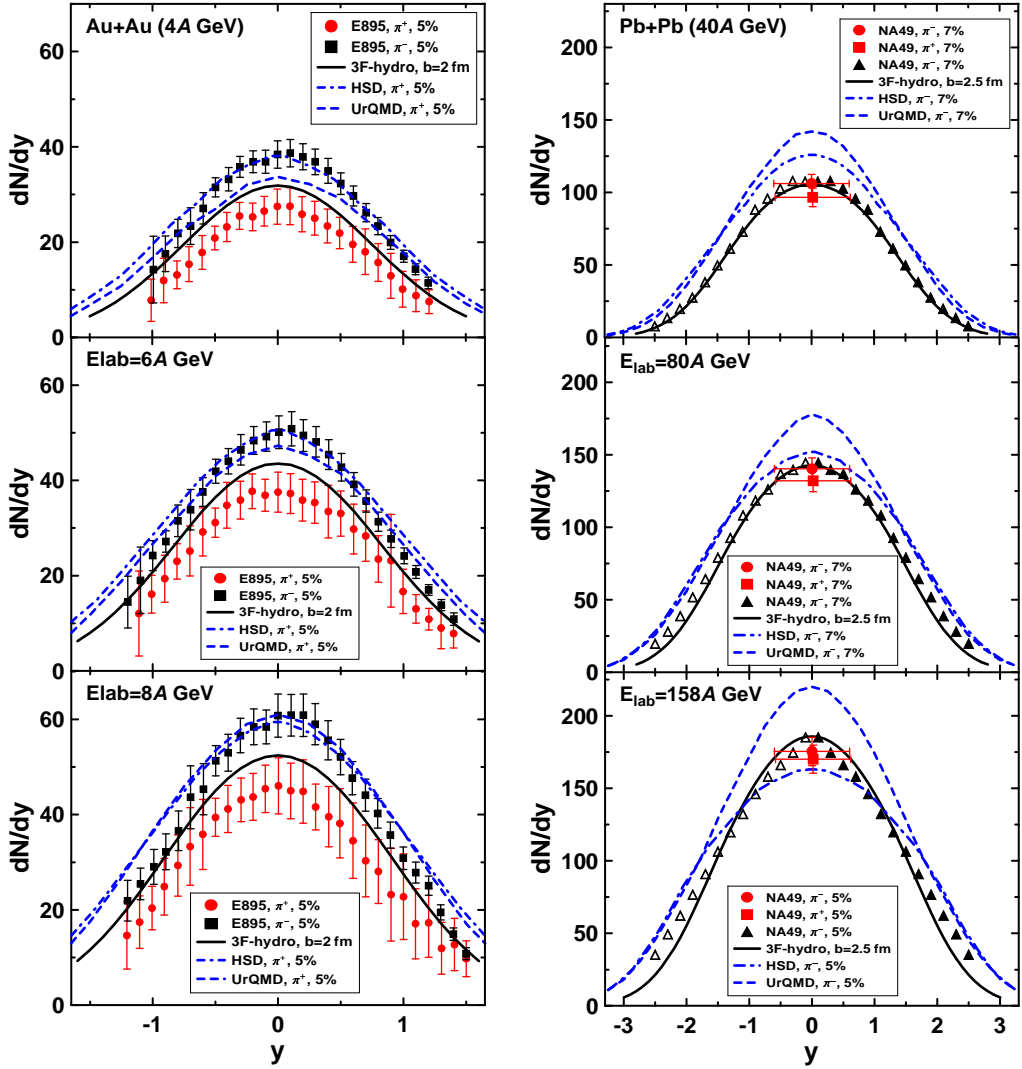}
\caption{(Color online) Pion rapidity spectra from central collisions
  at the AGS 
(left panel) and SPS (right panel) energies. Experimental points
are taken from~\cite{E895} (AGS energies) and \cite{NA49-pi} (SPS
energies). Notation is the same as in Fig. \ref{fig8}.}
 \label{fig9}
\end{figure*}
\begin{figure*}[t]
\includegraphics[height=14.cm]{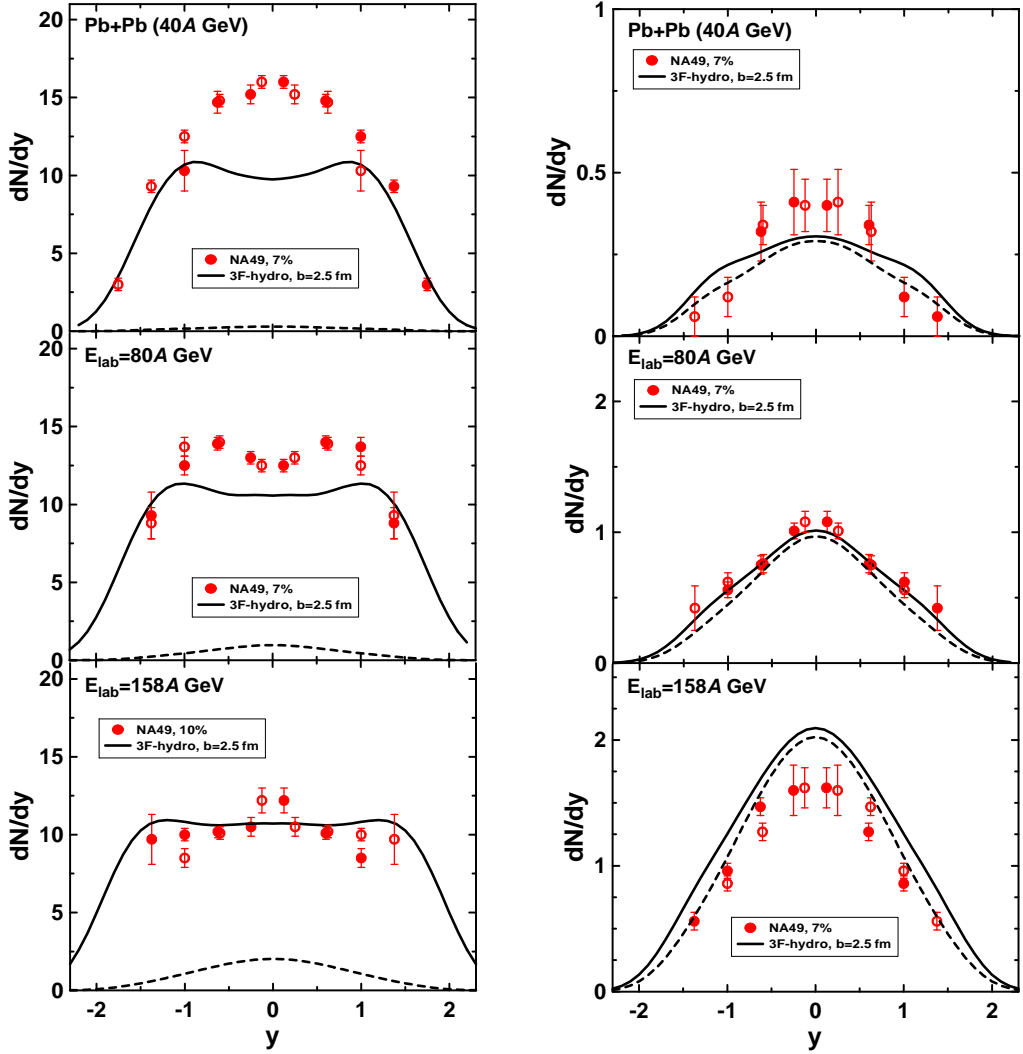}
\caption{(Color online) Rapidity spectra of
$\Lambda+\Sigma^0$ hyperons (left panel) and
$\bar{\Lambda}+\bar{\Sigma^0}$ antihyperons (right panel) 
from central ($b$ = 2.5 fm) Pb+Pb collisions. Solid
lines represent results of the 3-fluid 
model. Contributions from the fireball fluid are shown
by dashed lines. Preliminary experimental data are taken
from Ref. \cite{NA49-L}. The percentage indicates 
the fraction of the total reaction cross section, 
corresponding to experimentally selected events.}
 \label{fig10}
\end{figure*}
\begin{figure}[ht]
\includegraphics[width=7.9cm]{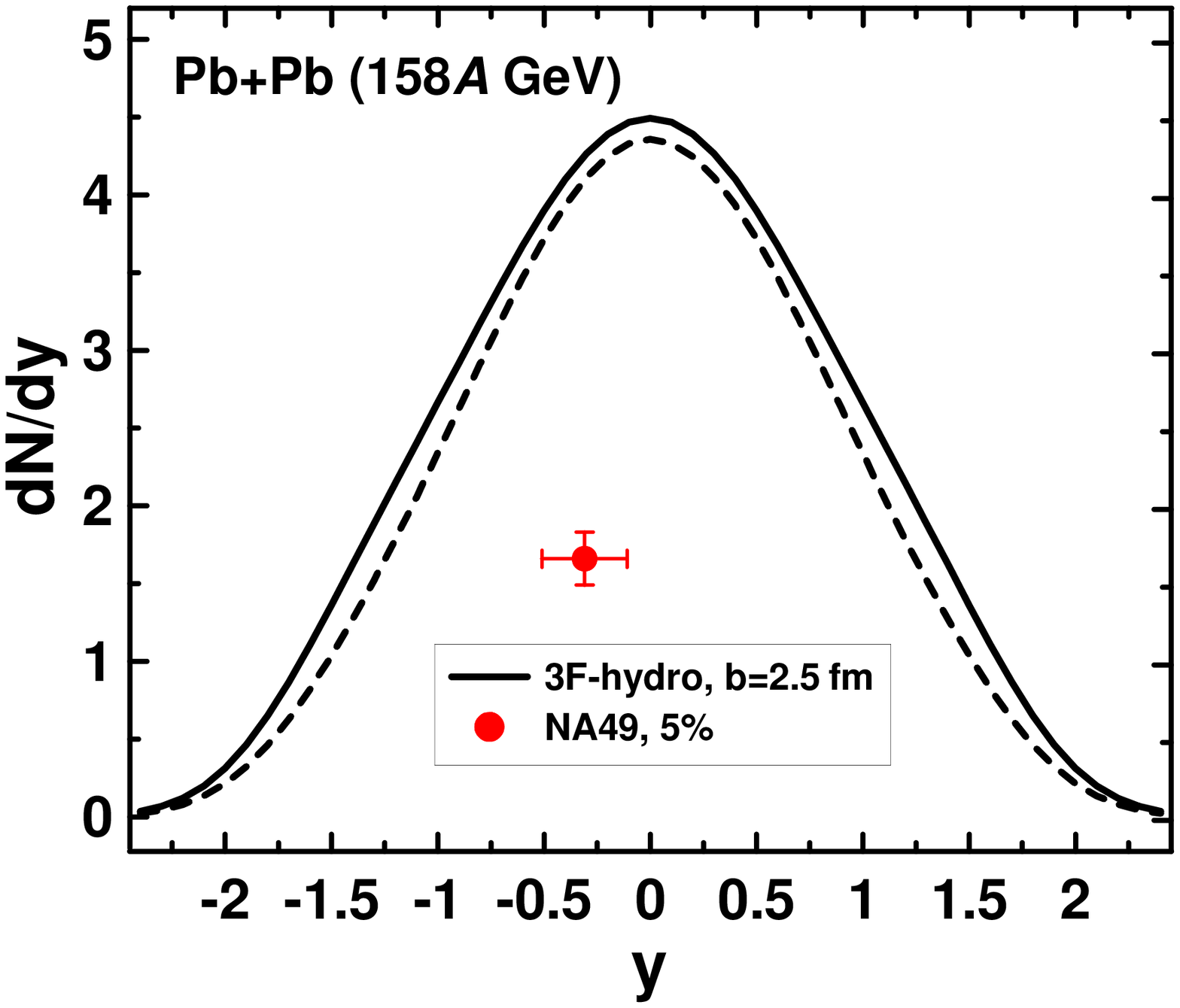}
 \caption{(Color online) Antiproton rapidity spectrum from central
   Pb(158$A$ GeV)+Pb  collisions. Notation is the same as in Fig. \ref{fig10}. 
Experimental data are taken from Ref. \cite{NA49-ap}.}
 \label{fig11}
\end{figure}

Once the friction has been already fitted to reproduce proton rapidity 
distributions, we can vary only the freeze-out energy density
$\varepsilon_{\scr{frz}}$ and the formation time of the fireball fluid
$\tau$. 
In contrast to rapidity distributions, 
the proton transverse-mass distributions turn out to be more
sensitive to $\varepsilon_{\scr{frz}}$, since their slopes reflect the
effective temperature of the frozen-out distributions. 
This is because the internal excitation certainly dominates in the 
transverse direction as compared to the collective motion. 
At the same
time, these  distributions are rather insensitive to
$\tau$, as well as all other baryonic (however, not antibaryonic) quantities. 
Therefore, to reproduce them, we choose
$\varepsilon_{\scr{frz}}\simeq$ 0.2 GeV/fm$^3$ at all incident energies 
with the only exception: $\varepsilon_{\scr{frz}}\simeq$ 0.1 GeV/fm$^3$
at $E_{\scr{lab}} \leq 2A$ GeV. 
As for low energies, $E_{\scr{lab}} \leq 2A$ GeV, 
we had to reduce $\varepsilon_{\scr{frz}}$, since otherwise 
the freeze-out would occur at the very early stage of the collision. 
The approximate nature of the $\varepsilon_{\scr{frz}}$
values results from the numerical
realization of the freeze-out procedure (cf. App. A).

Proton transverse-mass spectra 
at the AGS energies are exemplified in Fig. \ref{fig4}. They exhibit a
typical exponential fall-off. As seen, the 3-fluid 
model  well reproduces both the normalization and the slopes of this
fall-off, as well as their rapidity dependence. 
The
right panel of Fig. \ref{fig4} shows the dependence of the transverse mass
distributions at the mid rapidity on the impact parameter
$b$. As before, the values of impact
parameters were taken from the experimental estimate of
centrality of nuclear interaction \cite{E917}.
The impact-parameter dependence is reasonably reproduced by the model
as well even at relatively large impact parameters. 

The beam-energy dependence of the proton transverse-mass distributions
is presented in Fig. \ref{fig5} for central Pb+Pb collisions at 
the SPS energies. Overall, these distributions are in agreement
with experimental data. However, the 3-fluid 
model does not exhibit deviation from the exponential fall-off at
$(m_T-m)\lsim 0.2$ GeV, observed in experiment, most clearly at
$E_{\scr{lab}}=158A$ GeV. 
The same feature of the hydrodynamic calculation was earlier reported
in Ref. \cite{HS95}. As it was shown \cite{HS95}, a
post-hydro kinetic evolution is required to produce the observable two-slope
structure of the $m_T$-spectra. In our model such a post-hydro
evolution is absent. It is worthwhile to note that the above problem
is not an inalienable feature of any hydrodynamic calculation. For
instance, the deviation from the exponential fall-off was
reproduced in calculations by Kolb {\em et al.} \cite{1-hydro}.

\subsection{Pion Rapidity Distributions}
\label{Pion Rapidity}

Production of new particles is closely related to the amount of
entropy accumulated in the system. At the late stage of the collision,
the system expansion is isoentropic, i.e. the total entropy is
conserved. This is indeed so in our simulations, as we have checked
it. This fact implies that at high incident energies
$E_{\scr{lab}}\gsim 40A$ GeV the total number of produced pions
is approximately conserved during this expansion stage, since thermal 
pions are the dominant component among the produced particles. 
From the practical point of view, it means that the pion number is
approximately independent of the freeze-out energy density
$\varepsilon_{\scr{frz}}$. This is indeed observed in actual
simulations. However, at high incident energies the pion number depends
on the formation time of fireball fluid $\tau$, since 20$\div$30\%
of the pions are produced in the fireball fluid. The shorter
$\tau$ is, the earlier the fireball fluid starts to interact with baryonic 
subsystem, and hence the fewer pions survive in this fluid.
This implies that the formation time effectively 
tunes the strength of the fireball-projectile(target) friction which was 
only roughly estimated (cf. subsect. \ref{Interaction between 
Fireball and Baryon-Rich Fluids}).

At the first glance, the above speculation contradicts to the pion
absorption mechanism (cf. in Subsect. 
\ref{Interaction between Fireball and Baryon-Rich Fluids}), which
assumes that pions are simply captured by the baryon-rich fluids
without loosing their number. However, the pion number is not a
conserved quantity. In the baryon-rich fluids
the energy of the captured pions is thermally redistributed 
(accordingly to thermal chemical equilibrium) between kinetic energy
of (mainly) baryons and thermally produced (mainly) pions. Therefore,
their number is effectively reduced, since in the 
baryon-free fireball fluid the same energy was mainly accumulated in
thermal pions.

At lower incident energies $E_{\scr{lab}}\lsim 20A$ GeV, the
thermal pion production is not already so dominant but pions are
rather produced  through decays of resonances. 
Therefore, their number starts to depend on $\varepsilon_{\scr{frz}}$. 
The later freeze-out occurs, the fewer pions are produced.
At the same time, the
contribution of the produced fireball fluid becomes less important,
and hence the  $\tau$-dependence of the pion multiplicity becomes
weak.  At intermediate energies 
$20 \lsim E_{\scr{lab}}\lsim 40A$ GeV, there is a moderate
dependence of the pion multiplicity on both $\varepsilon_{\scr{frz}}$
and $\tau$. 
Note that these parameters,  $\varepsilon_{\scr{frz}}$ and $\tau$, 
mostly affect the overall 
normalization of pion rapidity spectra, whereas the shape of these
spectra mainly depend on the stopping power, which is kept fixed here.

As both the friction and $\varepsilon_{\scr{frz}}$
have been already fixed above, the pion rapidity distributions at low
incident energies $E_{\scr{lab}}\lsim 20A$ GeV can be 
considered as predictions of the model. In order to reproduce the pion
yield at high incident energies $E_{\scr{lab}}\gsim 30A$ GeV ,
the formation time of the fireball fluid has been taken $\tau=$ 2 fm/c. 
Note
that $\tau$ is the formation time of a fluid element, consisting
of a number of particles
(cf. Eq. (\ref{tau-part}) and discussion above it).
This implies that
the formation time of a separate particle (pion) at rest 
is certainly shorter
than $\tau$. Assuming that the temperature scale of the fireball fluid 
is approximately 100 MeV, we estimate that the formation time of a 
separate pion at rest approximately equals 1 fm/c. 
This is in good agreement with the estimate of the string-formation time. 
However, one should keep in mind that the latter value refers to 
a separate string in vacuum, i.e. without any effects of string 
interactions (junctions and color-rope formation). At the same time, 
our $\tau$ has an in-medium sense with all these multi-particle effects 
included. Therefore, their coincidence should be 
taken with caution. 
Note that with $\tau=0$ the multiplicity of pions (as well as other
newly produced particles)  would be 10--15\% underestimated at high
incident energies ($>30A$ GeV). 
All other observables would remain  approximately the same.

Since particles are not isotopically distinguished in our model, we 
assume that numbers of pions of each charge are  equal, i.e. 
$N_{\pi^+}=N_{\pi^-}=N_{\pi^0}=N_{\pi}/3$, where $N_{\pi}$ is the 
calculated total number of pions. Of course, this is a rough estimate 
of $\pi^+$ and $\pi^-$ yields, since $N_{\pi^-}$ always exceeds 
$N_{\pi^+}$ because of the initial isotopic asymmetry of colliding nuclei. 
Therefore, only if our calculation of $N_{\pi}/3$ complies with experimental 
$(N_{\pi^+}+N_{\pi^-})/2$, we refer to this as "a good agreement".

Fig. \ref{fig8} represents pion rapidity distributions in central
Au+Au collisions at the incident energy $E_{\scr{lab}}=$ 10.5$A$ GeV.
Sensitivity of this distribution to the variation of the impact parameter 
is also demonstrated. 
As seen,
the hydrodynamic results indeed fall in between the $\pi^+$ and $\pi^-$ 
data, hence they comply with these data. 
In contrast, the
$\pi^+$ rapidity spectra, calculated in two transport models, 
HSD and UrQMD models \cite{WBCS03} also displayed in Fig. \ref{fig8}, 
closely follow  experimental points but for
negative pions. 
The overestimate of the pion yield by the kinetic models
can be attributed to a lack of collective interaction in the 
EoS's corresponding to those models.

All above said is in fact true for the whole energy range under
consideration,   
as it is demonstrated in Fig. \ref{fig9}. The 3-fluid
model reasonably reproduces the pion distributions, while the 
HSD and UrQMD models certainly overestimate them.

\subsection{Rare Particle Production}
\label{Rare Particle}

Now all the parameters of the model are fixed. Therefore, all further 
calculations can be treated as predictions of the 3-fluid model.

As an example of the hydrodynamic description of rare channels,
the rapidity spectra of $\Lambda$ and $\bar \Lambda$ hyperons are
presented in Fig. \ref{fig10}. Contributions from decays of higher
resonances are taken into account. Both shape and absolute value
of hyperon and antihyperon spectra are reproduced surprisingly well.
The $\Lambda$ hyperons originate mainly from baryon-rich fluids and
shapes of their spectra closely follow those of the protons. 
Although the experimental data are preliminary,
one may note that agreement with experiment becomes certainly worse
at the incident energy $E_{\scr{lab}}=$ 40$A$ GeV. 
Unfortunately, the only measured point in proton rapidity
distribution (see Fig. \ref{fig3}) does not allow us to 
conclude on either similarity or difference of the $\Lambda$ and $p$ 
distributions at this energy. In
contrast, $\bar \Lambda$ antihyperons are dominantly created in the
baryon-free fireball and have typical single-bump thermal spectra.

At the same time the experimental data on antiproton production are
certainly overestimated in the our model, as
illustrated in Fig. \ref{fig11}. Since particles are not 
isotopically distinguished in our model, we have estimated the antiproton 
yield as half of that of antinucleon: 
$N_{\bar{p}} =  \frac{1}{2}N_{\bar{N}}$, assuming that $p\bar{p}$ and 
$n\bar{n}$ pairs are produced with approximately equal probability. 
The contributions of weak-decay channels $\bar{\Lambda} \to \bar{N}+\pi$ and 
$\bar{\Sigma} \to \bar{N}+\pi$ are excluded.


\begin{figure}[thb]
\includegraphics[width=8.4cm]{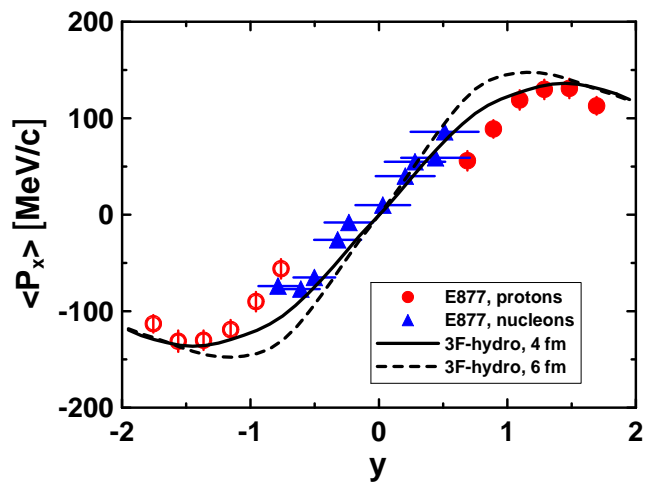}
\caption{(Color online) Transverse-momentum flow of nucleons and protons as a
  function of rapidity 
for semi-central Au+Au collisions at $E_{\scr{lab}}=$ 10.5$A$
GeV. The 3-fluid calculations are presented for $b=$ 4 fm 
(solid line) and $b=$ 6 fm (dashed line).  The data for protons (circles)
and all baryons (triangles) are taken from Ref. \cite{E877-px}.
Full symbols correspond to measured data, open ones are those reflected 
with respect to the mid rapidity point. } \label{fig6}
\end{figure}

\subsection{Flow}
\label{Flow}

Flow quantities of different types quantify space--momentum
correlations of collective motion of the strongly interacting matter. This
collective motion is essentially caused by the pressure gradients
arising during the evolution of the collision and hence is intimately 
related to the EoS and, in
particular, to a possible phase transitions. A spectacular loss of
correlation between the observed particle transverse momenta and
the reaction plane, which gives rise to dramatic reduction of the
directed flow, has been predicted by Rischke  et al. \cite{R96} in
the conventional hydrodynamic model with the bag-model EoS. The
subsequent studies showed that the observed signal essentially
depends not only on the EoS but also on the collision dynamics 
\cite{Brac00a,INNTS}.

The conventional transverse-momentum flow of an $a$ species is defined 
as \cite{DO85}, cf. Eq. (\ref{sp_y-mT}),
\begin{widetext}
\begin{eqnarray}
\langle p_x^{(a)}\rangle (y)= \frac{\displaystyle \int d^2 p_T \ p_x  
\left[ E \ dN_a/d^3p + 
\sum_{R} E \ d^3 N_a^{(R \to a+X)}/d^3 p\right]}%
{\displaystyle \int d^2 p_T 
\left[ E \ dN_a/d^3p + 
\sum_{R} E \ d^3 N_a^{(R \to a+X)}/d^3 p\right]}
, 
\label{eq-px}
\end{eqnarray}
\end{widetext}
where $p_x$ is the transverse momentum of a particle in the reaction
plane, and integration runs over the transverse momentum $p_T$. 
The second term in square brackets takes into account the contribution of 
resonance decays, resulting in $a$ production, cf. App. \ref{Resonance
  Decays}, the sum runs over all relevant resonances $R$.

As seen in Fig. \ref{fig6}, the 3-fluid model reasonably 
reproduces a  general trend of the baryonic $\langle p_x\rangle (y)$ 
distribution at $E_{\scr{lab}}=$ 10.5$A$ GeV, exhibiting two
clear peaks near the target and projectile rapidities. 
In fact, we compute $\langle p_x\rangle (y)$ of so-called
primordial nucleons, which later may coalesce, forming light
fragments. 
In view of this, it is not surprising that agreement with the nucleon 
$\langle p_x\rangle (y)$ data, which indeed take into account
contribution of light nuclear fragments,  
seems to be better than that with identified-proton data.

\begin{figure*}[thb]
\includegraphics[width=12.9cm]{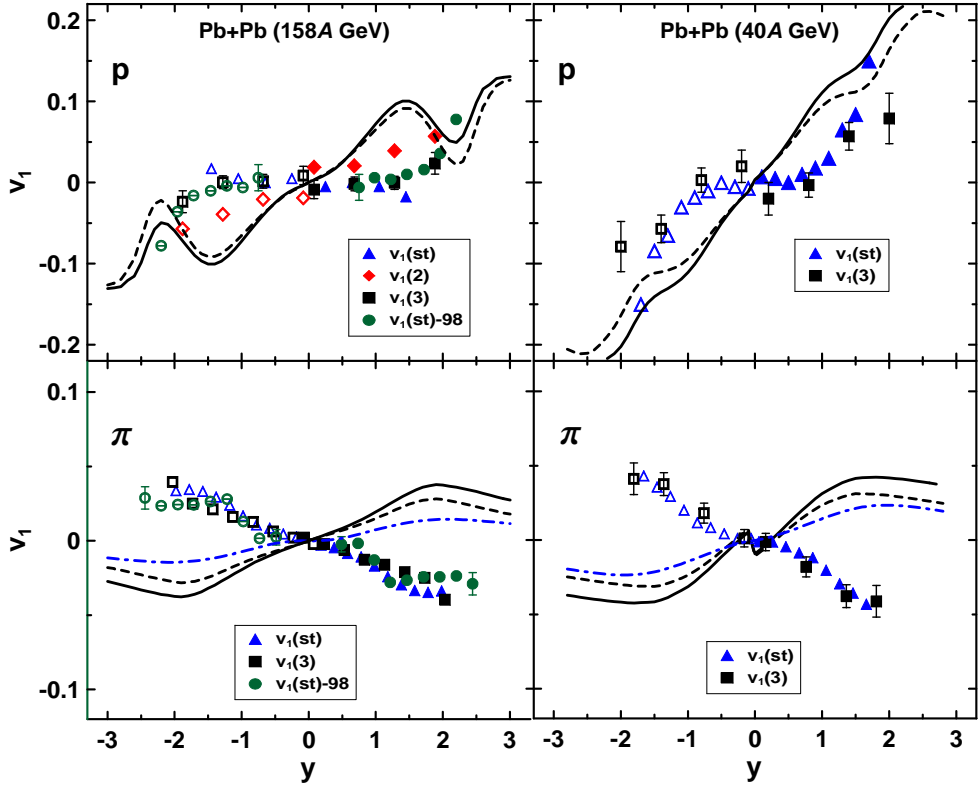}
\caption{(Color online) Directed
flows of protons (upper panels) and charged pions (lower panels) in 
semi-central Pb+Pb collisions at $E_{\scr{lab}}=$ 158$A$ (right panels) and 
40$A$ (left panels)  GeV as functions of rapidity. 
The 3-fluid calculations are presented 
for $b=$ 5.6 fm (solid line) and $b=$ 4 fm (dashed line).
The dot-dashed lines show contributions of
the fireball fluid. Experimental data are taken
from Ref. \cite{NA49-03-v1}. These data were obtained by two different
experimental procedures: the standard one ($v_{1,2}(st)$) 
and the method of $n$-particle correlations ($v_{1,2}(n)$). 
Full symbols correspond to
measured data, while the open symbols are those reflected  
with respect to the mid rapidity. Updated data of the NA49
Collaboration \cite{NA49-98-v1} ($v_{1,2}(st)-98$) with acceptance 
0.05 $<p_T<$ 0.35 GeV/c for pions  and
0.6 $<p_T<$ 2.0  Gev/c for protons, are also shown.}  
\label{fig7}
\end{figure*}
\begin{figure*}[thb]
\includegraphics[width=12.9cm]{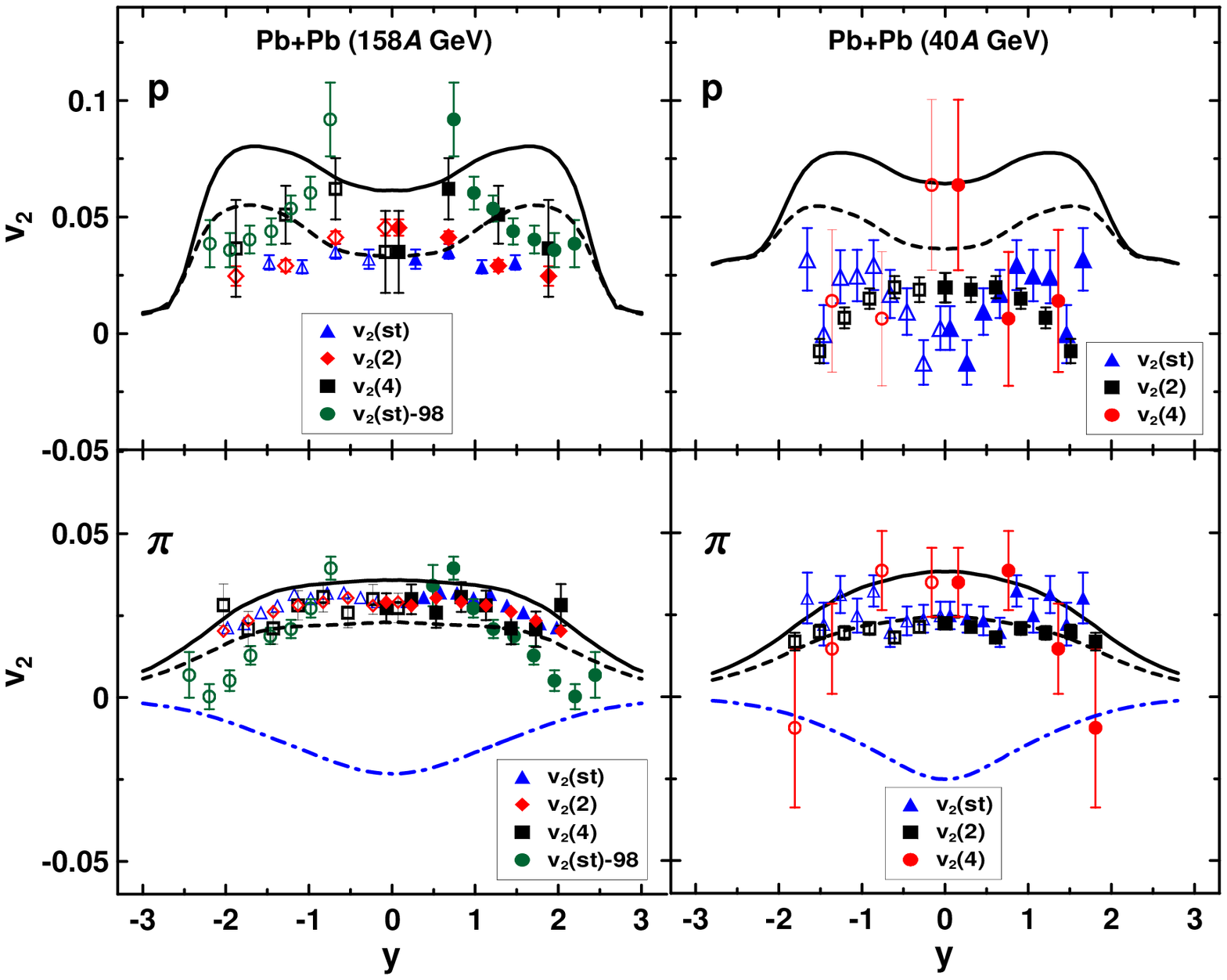}
\caption{(Color online) 
The same as in Fig. \ref{fig7} but for the elliptic flow.}
\label{fig7a}
\end{figure*}

At high energies the azimuthal asymmetry is usually characterized by
the first and second coefficients of Fourier expansion of the
azimuthal-angle dependence of the
 single-particle distribution function, i.e. by the directed flow
$v_1=\langle \cos \phi\rangle$ and elliptic flow  
$v_2=\langle \cos 2 \phi\rangle$: 
\begin{widetext}
\begin{eqnarray}
 \label{eq-v1}
v_1^{(a)} (y)&=& 
\frac{\displaystyle \int d^2 p_T  \left(p_x/p_T\right)
\left[ E \ dN_a/d^3p + 
\sum_{R} E \ d^3 N_a^{(R \to a+X)}/d^3 p\right]}%
{\displaystyle \int d^2 p_T 
\left[ E \ dN_a/d^3p + 
\sum_{R} E \ d^3 N_a^{(R \to a+X)}/d^3 p\right]}
, 
\\
v_2^{(a)} (y)&=& 
\frac{\displaystyle \int d^2 p_T \ \left[(p^2_x- p^2_y)/p^2_T\right]
\left[ E \ dN_a/d^3p + 
\sum_{R} E \ d^3 N_a^{(R \to a+X)}/d^3 p\right]}%
{\displaystyle \int d^2 p_T 
\left[ E \ dN_a/d^3p + 
\sum_{R} E \ d^3 N_a^{(R \to a+X)}/d^3 p\right]}
. 
\label{eq-v2}
\end{eqnarray}
\end{widetext}
Examples of the $v_1$ and $v_2$ flow for protons and pions at 
$E_{\scr{lab}}=$ 40$A$ and 158$A$ GeV are presented in Figs 
\ref{fig7} and \ref{fig7a}. The declared experimental percentage of
the total reaction cross section, related to the experimental
event selection,  allows us to estimate the
corresponding impact parameter as  $b=$ 5.6 fm. In view of uncertainty
of this estimate and in order to reveal the model dependence on the
impact parameter, we also show calculations with $b=$ 4 fm.  
Flows of the fireball fluid calculated at $b=$ 4 and 5.6 fm are hardly
distinguishable. 
Several sets of data, which noticeably
differ from each other, are shown in these figures. 
The "standard" method \cite{DO85,PV98} for evaluating the flow
 coefficients requires an event-by-event estimate of the reaction
 plane with which outgoing hadrons
  correlate. However, this method does not discriminate other sources of
 correlations, like those due to global momentum conservation,
 resonance decays, etc. 
 Recently a new method of
$n$-particle correlations has been proposed, which allows to get
 rid of these non-flow correlations in extracting $v_1$ and $v_2$ from
 genuine azimuthal correlations \cite{BDO01}. Note that only statistical 
 errors are indicated 
in these  figures. The systematic error for protons is 0.005 for
$v_2$ and 0.01 for $v_1$ at 158$A$ GeV,  they can be by 50$\%$
larger at 40$A$ GeV \cite{NA49-03-v1}.

At these energies the 3-fluid $v_1$ flow 
significantly differs from the data. 
Overall, the 3-fluid model predicts an 
essentially stronger directed flow than that experimentally observed. 
This disagreement can not be caused by neither uncertainties in
the impact parameters nor application of different methods for
measuring the directed flow \cite{NA49-03-v1}.
Moreover, the calculated pion directed flow closely follows the pattern of 
the proton one, while the pion $v_1$ data reveal anticorrelation with 
proton flow. Note that the $v_1$ flow in the fireball fluid is very weak. 
Therefore, it is not surprising that pion and proton $v_1$ are
correlated in the  
3-fluid model, since they reflect the same collective motion of the 
baryon-rich fluids.

The reasons for this poor reproduction of $v_1$ can be
two-fold. Certainly the first reason consists in disregarding the 
fact that a part of frozen-out
particles is ``shadowed'' by still hydrodynamically evolving matter
(cf. discussion in Subsect. \ref{Freeze-Out}). 
This shadowing means that frozen-out particles cannot 
freely propagate through the region still occupied by the hydrodynamically 
evolved matter but rather get reabsorbed into the hydrodynamic phase%
\footnote{
Of course, the conventional shadowing (of participants by spectators)
is automatically taken into account in our calculations: 
there simply is no emission from our participants 
into the direction of spectators, since they are not separated by any
freeze-out region. Moreover, the ``participants'' and ``spectators''
are very approximate terms for different parts of the unified
hydrodynamic system.
}. 
Apparently, the baryon directed flow is less affected by this 
shadowing. The reason is that 
the baryon directed flow reveals the collective flow of matter, which
is mainly built of baryons as the most abundant and heavy component of
the system. This collective flow is mainly formed at the early stage
of the reaction. Baryon rescatterings within this earlier-formed
collective flow at later (freeze-out) stages do not 
essentially alter the collective transverse momentum of the matter. 
At the same time, the pions can be much stronger
affected by this   
shadowing, since they are screened by the predominantly baryonic matter, 
where pions may be essentially decelerated or even absorbed. This can
drastically  change the pion $v_1$.

The second possible reason is 
very probable from our point of view. It could happen that the
hadron EoS used in these simulations is too hard, i.e. results in too strong
bounce-off of matter. A softer EoS, in particular due to possible
phase transition  
\cite{R96,Brac00a}, would reduce the strength of the proton directed flow in 
favor of its better reproduction. Then the problem of
correlation/anticorrelation  
with the pion flow would be more delicate, because the shadowing effect could 
easily change the sign of the pion flow. 
An additional argument in favor of a 
softer EoS is the success of microscopic RQMD~\cite{LPX99} and
UrQMD~\cite{BBSG99} 
models, which correspond to essentially softer EoS, since it is closer
to that of the gas. 
These models qualitatively reproduced early
measurements of $v_1(y)$ and  $v_2(y)$ for both protons and pions at
$E_{\scr{lab}}=158A$ GeV. Recently a good
microscopic description of the differential flow at 40$A$ GeV
was also obtained~\cite{SBBSZ04}.

The calculated elliptic flow of protons (see Fig. \ref{fig7a}) is
positive ("in plane") and somewhat overestimates experimental points
at both incident energies, 40$A$ and 158$A$ GeV. Though data at 40$A$
GeV are still controversial, this overestimate 
again indicates that the used hadronic EoS is too hard. 
At the same time, the hydrodynamic pion
elliptic flow is in surprisingly good agreement with experimental 
data of the NA49 Collaboration both in magnitude and shape, exhibiting 
a shallow minimum at the mid rapidity. 
The elliptic flow  $v_2$ results from 
the initial spatial asymmetry of non-central nucleus-nucleus
collisions. The overlap  lens-shaped geometry of two nearly
thermalized nuclei is reproduced correctly in the 3-fluid
model. As this lens-shaped matter expands, it produces the elliptic
flow. Note that mesons emitted from the
fireball fluid have negative ("out of plane") $v_2$ flow.

The earlier pion $v_2$ data of the NA49 collaboration, taken at smaller
acceptance at the top SPS energy \cite{NA49-98-v1}, have been already 
analyzed within hydrodynamic approaches.    
In the expansion model \cite{KSH99} the Bjorken scaling
solution \cite{B83} was assumed for longitudinal evolution and 2D hydro was
solved numerically for transverse one. In this way, the elliptic flow
could be estimated only at the mid rapidity point. 
The full 3D expansion
model with postulated initial conditions was applied to the meson elliptic
flow by Hirano \cite{H01}. In a qualitative agreement with
Ref. \cite{KSH99}, it was found that $\rho$-meson decays result in 
almost vanishing azimuthal
anisotropy of pions near the mid rapidity. Let us remind that decays of all
relevant resonances are taken into account in our model. 

Recently, collective flows in heavy ion collisions from AGS to
  SPS energies were  systematically  studied 
in a transport model with various assumptions 
on the nuclear mean field \cite{IOSN05}. It was found that momentum dependence 
in the nuclear mean field is of prime importance for the reproduction
  of the $v_1$ 
and $v_2$ flows. It turned out that generally our  results are rather close 
to those of Ref. \cite{IOSN05}, if the mean-field momentum dependence
  is absent. This is precisely the case in our EoS. 
It is of interest to note that similarly to our results but in contrast with 
the experiment, the pion directed flow at $E_{\scr{lab}}=158A$  GeV
correlates 
with the proton $v_1$ independently of the version chosen for the mean 
field \cite{IOSN05}.

\begin{figure*}[ht]
\includegraphics[width=17.cm]{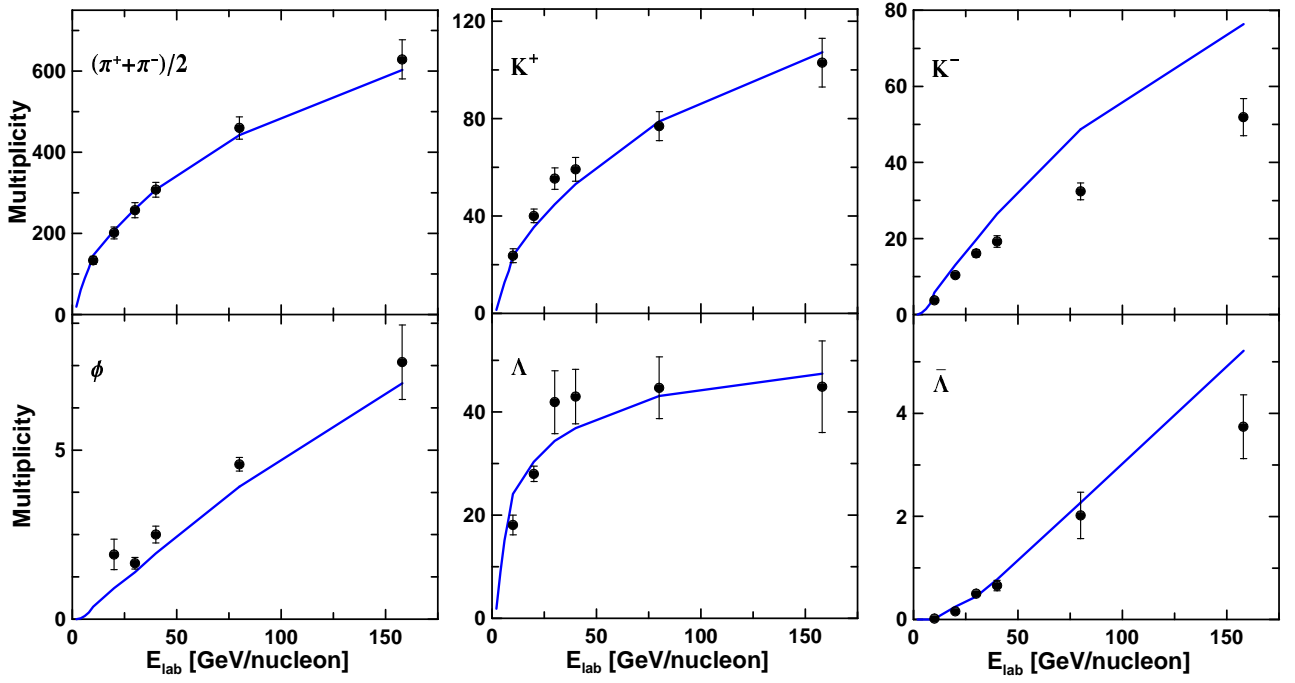}
\caption{(Color online) Multiplicities for central Au+Au
(at AGS energies) and  Pb+Pb (at SPS energies) collisions 
as functions of the incident energy. Open triangles connected by solid
lines represent the 3-fluid results. 
The experimental data (full circles) are taken
from Refs. 
\cite{NA49-pi,Mischke:2002ub,Anticic:2003ux,Fr:01,xi,pl,%
Alt:2004kq,Alt:2003rn,mg:04,cm:04,ar:04,cb:04} 
for SPS energies and from Refs. 
\cite{E802:1,E802:2,E802:3,Albergo02,Ahmad96} for AGS energies.}
\label{figM}
\end{figure*}

\begin{figure*}[htb]
\includegraphics[width=14cm]{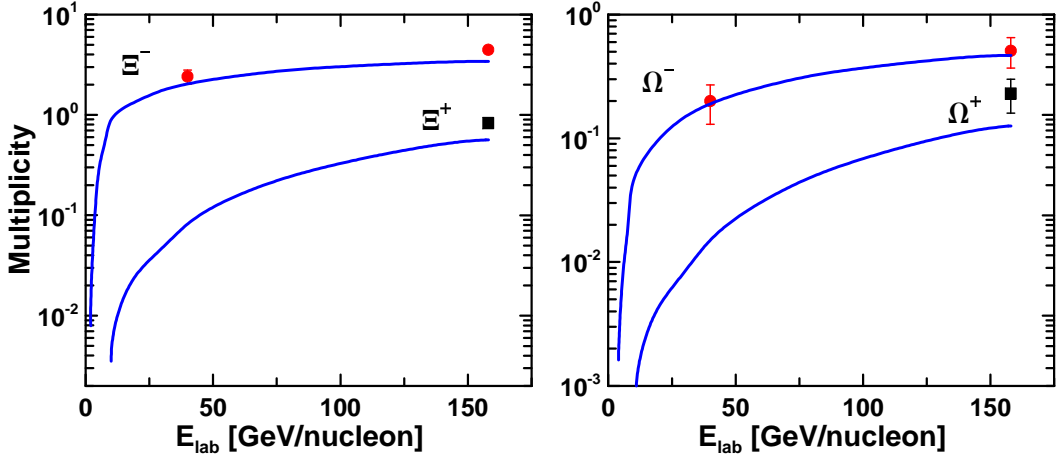}
\caption{(Color online) 
The same as in Fig. \ref{figM} but for multiplicities of
  multi-strange hyperons. The experimental
  data are taken from Refs. \cite{Alt:2004kq,NA49_Xi1,NA49_Xi2}: 
full circles for hyperons, and full squares for antihyperons. 
}  
\label{figMms}
\end{figure*}

\subsection{Multiplicities} 
\label{Multiplicities}

Though the quality of reproduction of particle multiplicities was
already clear from the above 
presented rapidity distributions, we would like to summarize them in
this subsection, as well as present those for hadrons not mentioned
above. In Fig. \ref{figM}, particle multiplicities for
central Au+Au 
(at AGS energies) and  Pb+Pb (at SPS energies) collisions 
as functions of the incident energy are presented and
confronted with the experimental data. The experimental data are taken
from Refs. 
\cite{NA49-pi,Mischke:2002ub,Anticic:2003ux,Fr:01,xi,pl,Alt:2004kq,%
Alt:2003rn,mg:04,cm:04,ar:04,cb:04} for SPS energies and from Refs. 
\cite{E802:1,E802:2,E802:3,Albergo02,Ahmad96} for AGS energies. These
data are presented as 
they were summarized in Ref. \cite{Becattini}, where they were scaled
to the same  5\% experimental trigger. Because of this reduction to
the same  5\%  trigger, all the simulations were performed at the same
central impact parameters corresponding to this trigger: $b=$ 2 fm for
Au+Au and $b=$ 2.5 fm for Pb+Pb. 
Corrections for feed down from
weak decays (mostly $\Xi^-$, $\Xi^0$ and their antiparticles)  were
not applied in the original 
published data. They  were, however,
estimated to be 6\% for $\Lambda$ and 
12\% for $\overline{\Lambda}$ \cite{Anticic:2003ux}. 
Based on this estimate the $\Lambda$ and $\bar{\Lambda}$
yields given in Ref. \cite{Becattini} were reduced by 6\% and 12\%,
respectively. This also complies with conditions of our simulations.

We compare our calculated pion multiplicities with the experimental
results for half-sum of those for $\pi^+$ and $\pi^-$, since the model
does not distinguish the isospin of particles. 
Note that the pion multiplicities were fitted by means of the
freeze-out parameter and the formation time of the fireball fluid,
while all other multiplicities are already predictions
of the model. As seen, the model reasonably reproduces the energy
dependence of various multiplicities, except  that for $K^-$. The latter
is strange, especially in view of that even the rare-channel particles,
such as $\phi$ and $\bar{\Lambda}$, are reasonably reproduced without
any additional tuning. 
Probably, a kind of post-hydro kinetic evolution, similar to that
performed in Ref. \cite{HS95}, is required for a proper reproduction
of the $K^-$ multiplicity. Indeed, cross sections of the reactions
$\bar{K}N\to\pi\Lambda$ and $\bar{K}N\to\pi\Sigma$ are very high at
low relative momenta \cite{CB}. 
The first reaction transforms a part of $K^-$ into $\Lambda$. The resulting
$\Lambda$ enrichment would not contradict to data, since the NA57
data for the $\Lambda$ production \cite{Antinori} are certainly above
the NA49 data presented in Fig. \ref{figM}. 
The second reaction results in a loss
of strangeness because of the weak decay $\Sigma^{\pm}\to N\pi$. 
In fact, this is the
dominant channel for the $\bar{K}$ absorption on nucleons at low
energies. 
In our model such a post-hydro evolution is absent. 
It is worthwhile to note that the statistical-model fit of the data
also overestimate $K^-$ multiplicities at the SPS energies
\cite{Andronic}.

At comparatively low incident energies $E_{\scr{lab}}<$ 10$A$ GeV, the
grand canonical treatment of the strangeness production, used in the
3-fluid model, becomes poorly applicable. Therefore, we do not
present the 3-fluid predictions at these energies. 
At energies $E_{\scr{lab}}=$ (30--40)$A$ GeV the observed
multiplicities of  $K^+$ and $\Lambda$ are certainly above the smooth curve
predicted by the 3-fluid model. This correlates with already 
above observed fact (see Figs \ref{fig10} and \ref{fig7a}) that
agreement with data in this energy region is certainly worse than at
other energies.

The excitation functions for production of multi-strange hyperons, 
$\Xi$ and  $\Omega$, and the corresponding antihyperons are presented in
Fig. \ref{figMms}. In Ref. \cite{Alt:2004kq}, $\Omega^-$ and $\Omega^+$ were
not separated at $E_{\scr{lab}}=40A$ GeV. Due to that their
total yield is plotted as $\Omega^-$, since the yield of $\Omega^+$
antihyperons is much lower than that of $\Omega^-$. 
It is seen that calculated multiplicities of multi-strange
hyperons are also in a reasonable  agreement with the available
data. Note that these yields are strongly underpredicted in the UrQMD
model \cite{Alt:2004kq}.

\section{Global Evolution of Nuclear Collisions}
\label{Global}

\begin{figure}[ht]
\includegraphics[width=8.5cm]{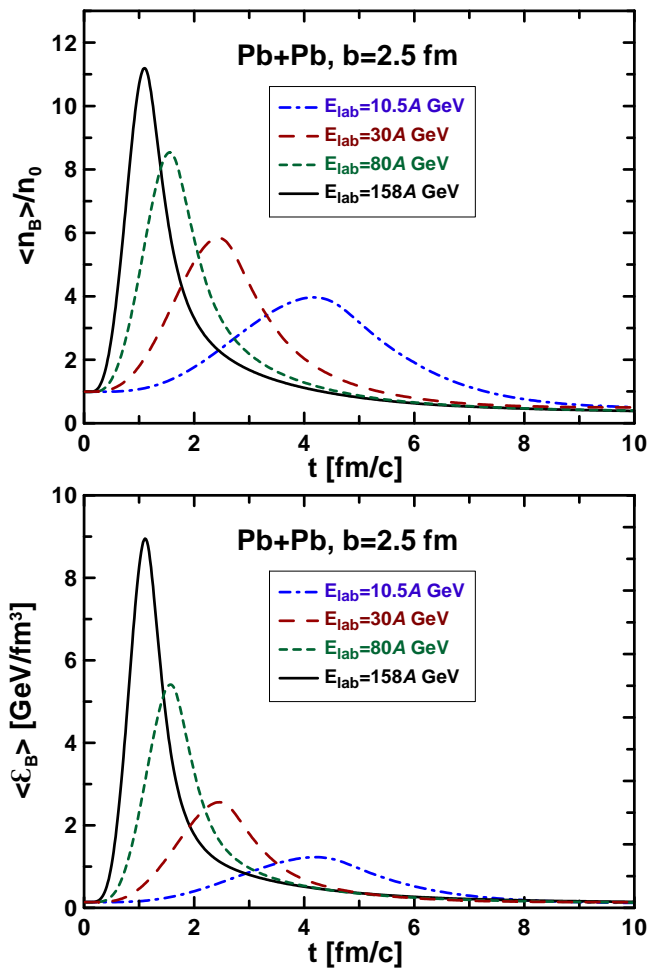}
\caption{(Color online) 
Temporal evolution of average baryon (upper panel) and energy
  (lower panel) densities 
for central Pb+Pb collisions within 3-fluid model. The time
is counted in the c.m. frame of the colliding nuclei. }
\label{fig12}
\end{figure}
\begin{figure}[ht]
\includegraphics[width=8.5cm]{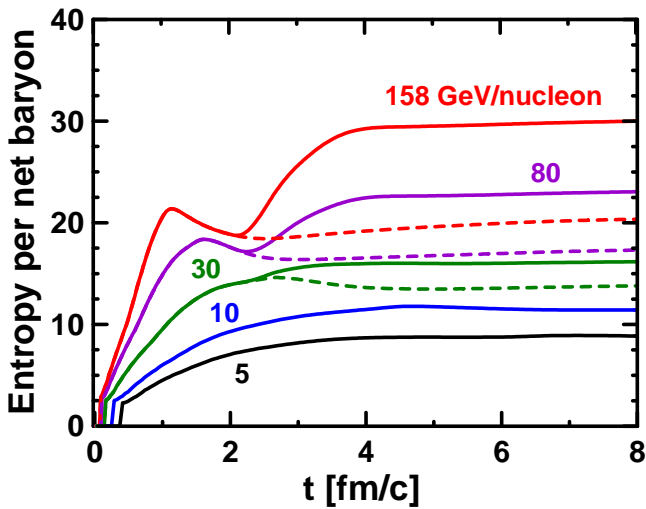}
\caption{
Time evolution of the 
entropy per baryon number in Pb+Pb central ($b$ = 2.5 fm)
collisions at various incident energies $E_{\scr{lab}}$. The total
entropy is displayed by solid lines, while the entropy of baryon-rich
subsystem -- by dashed lines. 
}
\label{fig-S}
\end{figure}

In the preceding Section, it was shown that the 3-fluid hydrodynamic
model with 
the simplest, purely hadronic EoS is able to reasonably
reproduce a great body of experimental data on relativistic heavy-ion
collisions. The considered observables 
characterize a state of the colliding system at the freeze-out stage.
With thus fixed parameters of the model, it
is possible to address the question of global evolution of 
nuclear collisions, e.g., the values of baryon and
energy densities achieved in the course of them. An important question
also is how long and in which volume these achieved values survive. 
 Dynamics
of heavy ion collisions at different bombarding energies is
illustrated in Fig. \ref{fig12}.
The shown baryon and energy densities are averaged over the whole
space occupied by fluids
\begin{eqnarray}
\label{avn}
\langle n_B\rangle&=& \int d^3x \ n_B \ W(x) /  \int d^3x \ W(x)~, \\
\langle \varepsilon\rangle&=& \int d^3x \ \varepsilon \ W(x) /  \int
d^3x \ W(x)~, 
\label{ave}
\end{eqnarray}
where $\varepsilon$ and $n_B$ are the proper (i.e. in the local rest
frame) densities, and 
the weight function is taken equal to the local proper baryon density
 $ W(x)=n_B(x)$. 
This weight function was chosen because we are primarily interested
here in evolution of dense baryonic matter. 
The 3-fluid nonequilibrium is quite strong at the initial stage of the
collision. 
The mean values of $\varepsilon$ and $n_B$ are calculated by means of 
averaging these quantities corresponding to either separate fluids (if
their mutual stopping has not occurred) or the unified fluid (if the
full stopping has happened). To characterize the degree of stopping, we used
 the auxiliary function introduced in Eq. (\ref{alpha}), which reveals
 quite a sharp transition between transparency ($\vartheta\simeq 1$) and
 full stopping ($\vartheta\simeq 0$). This convention
has been chosen in order to map the nonequilibrium configuration in
terms of equilibrium quantities, avoiding unphysical contributions of
the Lorentz contraction and collective motion into 
quantities  $\langle \varepsilon\rangle$ and $\langle n_B\rangle$. 
However, these results at early stages of collisions should be
taken with some care because of certain ambiguity of 
this mapping convention.

As seen from Fig. \ref{fig12}, the average baryon density $\langle
n_B\rangle$  increases quickly reaching a 
maximum, and then the expansion stage comes. The 
maximal compression of baryonic matter correlates with maximal
overlap of colliding nuclei. Evolution of the average energy density
$\langle \varepsilon\rangle$ proceeds in a similar way. Maximal values
of baryon 
and energy densities are high enough, exceeding hypothetic threshold
for deconfinement phase transition $\varepsilon \sim 1 \ GeV/fm^3$ 
even at $E_{\scr{lab}} \sim$ 10$A$ GeV. However, these values are slightly lower than those
reported in~\cite{Friman98}, because here  averaging over the whole
volume of colliding system is carried out rather than over
some fixed central region. Note that local
peak values may exceed these average values by a factor of about 2.

To get notion of the extent of thermalization achieved in the
collision process, in Fig. \ref{fig-S} we show the time evolution of the
entropy per baryon number (the latter is conserved quantity) in central
collisions. These calculations were performed without freeze-out in
order to keep account of the total entropy. We
have taken into account the entropy of only formed fireball fluid,
since for the unformed part even its definition is
ambiguous. Therefore, we deal with an open system, when not all the
fireball matter is formed. 
At the early stage of the collision (till the time instant of complete
overlap of nuclei) the entropy quickly rises, which
is the evidence of nonequilibrium in the system. After that the total
entropy either flatten (at $E_{\scr{lab}}<$ 40$A$ GeV)
or even slightly drops down (at $E_{\scr{lab}}>$ 40$A$ GeV). 
This entropy flattening naturally means thermalization. The temporal dropping
down of the entropy is related to still continuing production of
unformed fireball matter. It starts to form only later, when the
entropy starts to rise after the temporal fall. Therefore, at the
early stage of the reaction we deal with an open system. 
The entropy production in the baryon-rich fluids is proportional to
the term 
$$R=\left(u_{\scr p}\cdot u_{\scr t}\right)
\left(D_P-D_E\right)-\left(D_P+D_E\right),$$
if the fireball fluid is still unformed, cf. Eqs. (\ref{eq8p}),
(\ref{eq8t}) and (\ref{eq16}). This result follows from the
standard derivation \cite{Land-Lif} of hydrodynamic equation for the
entropy flow. Therefore, the fall or rise of the entropy of the
baryon-rich fluids depend on the sign of this $R$.
If $R>0$, the entropy rises. 
This occurs either at
$D_E=0$, since $\left(u_{\scr p}\cdot u_{\scr t}\right)\geq 1$,  
or when $\left(u_{\scr p}\cdot u_{\scr t}\right)\gg 1$, as it is the
case at the initial interpenetration stage. In the region, where the
entropy drops down, $D_E$ is still nonzero and 
$\left(u_{\scr p}\cdot u_{\scr t}\right)$ is not high enough,
therefore $R< 0$. In physical terms, the entropy drops down, when the
energy-momentum radiation into the fireball fluid 
$\propto -\left[\left(u_{\scr p}\cdot u_{\scr t}\right)+1\right]D_E$
dominates over the internal heating 
$\propto \left[\left(u_{\scr p}\cdot u_{\scr t}\right)-1\right]D_P$
of the baryon-rich subsystem.

\begin{figure}[hb]
\begin{center}
\includegraphics[width=8.5cm]{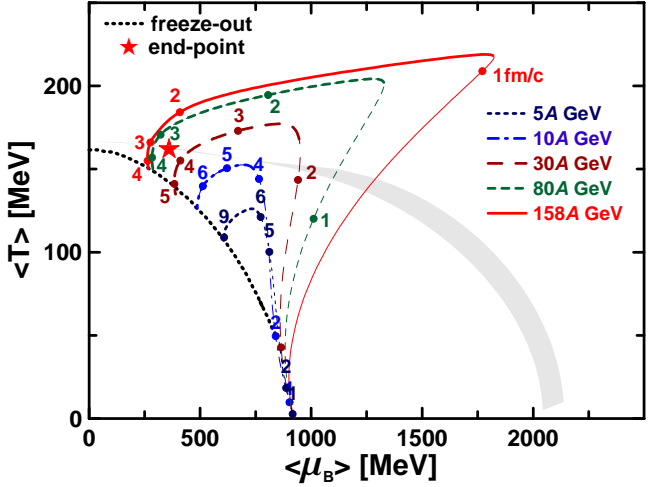}
\caption{(Color online) Dynamical trajectories in the ($T,\mu_B$) plane for
  central Pb+Pb collisions ($b$ = 2.5 fm) at various incident energies. 
Numbers near the dynamical
trajectories indicate the evolution time instants in the c.m. frame of
the colliding nuclei. 
Bold parts of trajectories are related 
to approximately thermalized
baryon-rich subsystem, 
while the thin ones -- to yet nonequilibrium evolution. 
The light-grey shaded region corresponds to the 
 boundary of the phase transition from the hadronic phase to the QGP, 
as it was estimated in Ref. \cite{TNFNR04}. 
Dotted line is the ``experimental'' freeze-out curve fitted to
observed multiplicities in the
approximation of the ideal gas model \cite{CR98} under condition
that the energy per hadron is 1 GeV. 
The star-symbol is the critical end-point calculated in
Ref.~\cite{Fodor01}. } \label{fig14} 
\end{center}
\end{figure}

As seen from Fig. \ref{fig-S}, at $E_{\scr{lab}}<$ 40$A$ GeV the
entropy is approximately conserved already after the colliding nuclei
overlap. These overlaps correspond to peak values in Fig. \ref{fig12}
and turning points of the
trajectories in Fig. \ref{fig14}. This fact suggests that
thermalization in these collisions occurs comparatively early. 
At higher incident energies, $E_{\scr{lab}}>$ 40$A$ GeV, 
the complete thermalization happens at the comparatively
late stage ($t\gsim 4$ fm/c), when the fireball fluid gets formed and
the entropy approaches the plateau. However, the entropy of the
baryon-rich fluids, displayed by dashed lines in Fig. \ref{fig-S},
slowly changes already after the complete overlap of
nuclei. It remains constant within 10\%. This baryon-rich subsystem 
can be considered as an approximately thermalized fluid still 
interacting with a ``bath'' of the fireball matter. 
Therefore, this overlap time can be approximately taken as a
time of  equilibration in the baryon-rich subsystem at all
considered here incident energies. 

%
\begin{figure}[thb]
\begin{center}
\includegraphics[width=8.cm]{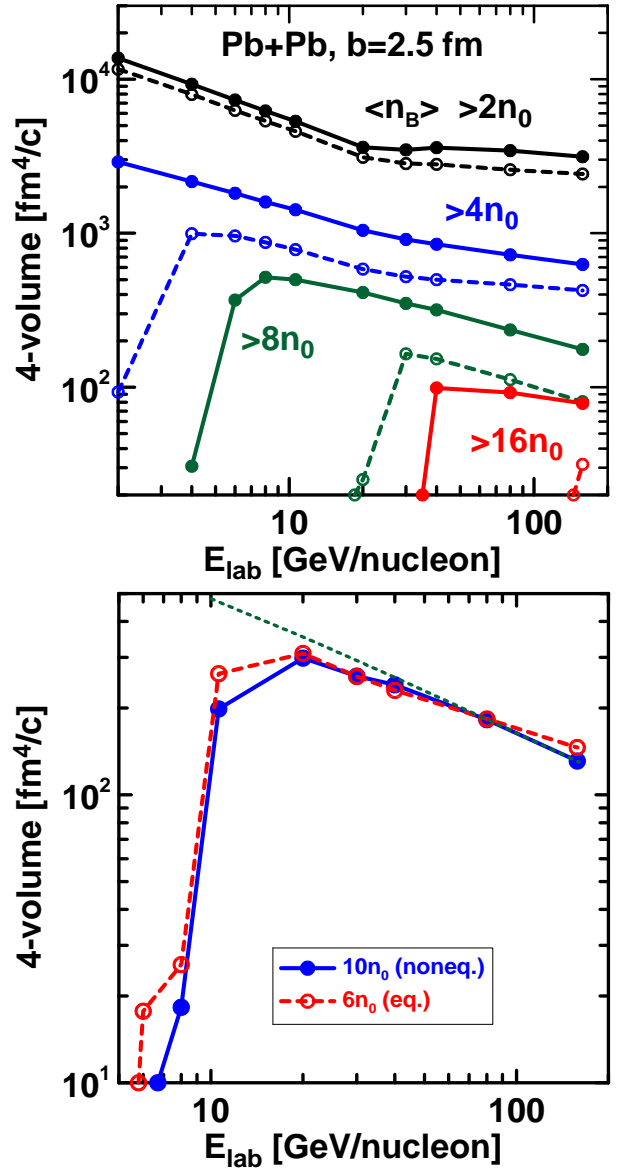}
\caption{(Color online) Invariant 4-volume corresponding to conditions on 
$n_B^{\scr{(noneq.)}}$ (solid lines) and on $n_B^{\scr{(eq.)}}$
(long-dashed lines) for central
Pb+Pb collisions as function of the incident energy. 
The dotted line on the lower panel is
the 4-volume $\Delta t\cdot\pi R^2\cdot
2R/\gamma_{cm}$ with $R=4$ fm, $\Delta t= 3$ fm/c (see the text).  
 }
\label{fig13}
\end{center}
\end{figure}

Based on the EoS used in simulations, the $n_B$ and
$\varepsilon$ densities can be recalculated in the
temperature--chemical-potential $(T,\mu_B)$ representation and
displayed in the conventional way as dynamic
trajectories in the ($T,\mu_B$) plane, see Fig. \ref{fig14}.
In view of the above discussion, we display approximately equilibrium
(for the baryon-rich subsystem) 
parts of the trajectories by bold lines, while those yet
nonequilibrium -- by thin lines, just in order to keep track of the
system evolution. 
As seen, the compression stage is rather short. For
example, at $E_{\scr{lab}}=$ 158$A$ GeV the  maximal values of $\langle
T\rangle$ and $\langle \mu_B\rangle$
are reached during the time of about 1 fm/c. After that a 
comparatively fast expansion, accompanied by rather slow
cooling, starts. At the incident energy of 
$\approx $ 30$A$ GeV, the  dynamical trajectory passes near the
critical end-point \cite{Fodor01}, where the order of the
deconfinement phase transition is changed. In fact, the trajectories
in the ($T,\mu_B$) plane 
strongly depend on the used EoS, even if they originate from the same
($n_B,\varepsilon$) initial state. Therefore, this should be
taken only as a rough estimate of the relevant incident energy, since
the  EoS of the hydro simulations has not involved any QGP phase
transition at all. 
The boundary of the phase transition from the hadronic phase to the QGP, 
estimated in Ref. \cite{TNFNR04} (light-grey shaded region
Fig. \ref{fig14}), is displayed just to remind on what may happen 
in this thermodynamic domain, if a EoS involves the phase transition. 
However, this trajectory representation is not
quite informative due to the following reason. 
In addition to the uncertainties of mapping of
nonequilibrium on equilibrium above discussed, 
the displayed time instants correspond to the cm frame of colliding
nuclei and hence are noninvariant. Therefore, Fig.~\ref{fig14} does not
tell us (in invariant way) how long the dense matter survives and in
which volume.

The way to overcome the above difficulties was proposed
in \cite{Friman98}. It consists in calculation of an invariant
4-volume $V_4$ in which a quantity $q$ exceeds a given value $Q$ 
\begin{eqnarray}
\label{V4-all}
V_4 (Q) = \int d^4 x \;\Theta (q-Q). 
\end{eqnarray}
This quantity provides a  Lorentz invariant measure of the
space--time region, where the quantity $q$ keeps high value $q\geq
Q$. In Fig. \ref{fig13} this 4-volume is shown for two cases:
$q=n_B^{\scr{(eq.)}}$ and  $q=n_B^{\scr{(noneq.)}}$. The first
  case corresponds to the  4-volume summed only over those regions, 
where full stopping (i.e. $\vartheta\simeq 0$ in terms of
Eq. (\ref{alpha})) has occurred and thermalized  $n_B$ 
exceeds certain value, while in the second case  the  4-volume is
summed over all regions (i.e. similarly to that in Eq. (\ref{avn})),
including those where stopping has not occurred. 
As seen, for production of the matter with comparatively high
densities ($n_B>4n_0$), there are certain 
preferable incident energies, which allow to attain the largest
4-volume. In particular, 
incident energies (10--40)$A$ GeV, planned at the new GSI
facility, are favorable for production of
equilibrated matter with baryon densities higher than $6n_0$. 
This conjecture is heavily based on the hadronic EoS used. However, we
expect that it is not too far from the truth, because we fairly well
reproduced the observed stopping power, which is of prime relevance to
the achieved compression.  
To get an impression of the scale of the corresponding 4-volume, 
we draw a short-dashed line presenting 
the 4-volume $\Delta t\cdot\pi R^2\cdot
2R/\gamma_{cm}$ composed of Lorentz-contracted cylindrical space
volume of radius $R=4$ fm and time interval $\Delta t= 3$ fm/c, here
$\gamma_{cm}$ is the $\gamma$-factor of
colliding nuclei in their c.m. frame. It is of interest that the
fall-off of the invariant $V_4$ is defined by this $\gamma$-factor.

\section{Discussion and Conclusions}

In this paper we have presented an extension of the 3-fluid model for
simulating heavy-ion collisions in the  range of incident energies
between few and about 200$A$ GeV. In addition to two
baryon-rich fluids, which constitute the 2-fluid model,  the
baryon-free fireball fluid of newly produced particles with delayed
evolution  is incorporated. This
delay is governed by a formation time, during which the fireball
fluid neither thermalizes nor interacts with the baryon-rich
fluids. After the formation, it thermalizes and comes into
interaction with the baryon-rich fluids. This interaction is
estimated from elementary pion-nucleon cross-sections.

The hydrodynamic treatment of heavy-ion collisions is an
alternative to kinetic simulations. The hydrodynamic approach has
certain advantages and disadvantages. Lacking the microscopic
feature of kinetic simulations, it overcomes their basic
assumption, i.e. the assumption of binary collisions, which is
quite unrealistic in dense matter. It directly addresses the
nuclear EoS that is of prime interest in heavy-ion research.
Naturally,
we have to pay for these pleasant features of hydrodynamics:
the treatment assumes
 that the non-equilibrium stage of the collision can be
described by the 3-fluid approximation. 

Taking advantage of the modern computers, substantial work has
been also done on improvement of numerics of the model. In particular,
it was found
that the numerical diffusion of the computation scheme should be as much
as possible avoided in order to get reliable results in hydro
simulations. To avoid the numerical diffusion the parameters of the
scheme should be carefully optimized, as it is described in
App. \ref{Numerics}.

The main unknowns of the present simulations can be briefly
summarized as follows: the equation of state and ``cross sections''.
The EoS is an external input to the calculation and thus can be
varied. Our goal is to find an EoS which in the best way reproduces
{\em the largest body of  available observables}. The ``cross sections'' are
equally important. They determine friction forces between fluids
and hence the nuclear stopping power. 
At present we
have at our disposal only a rough estimate of the friction forces
(cf. Ref. \cite{Sat90} for the friction of baryon-rich fluids and Sect.
\ref{Interaction between Fireball and Baryon-Rich Fluids} for
the friction with the fireball). 
Therefore, we have to fit the
friction forces to the stopping power observed in proton rapidity
distributions.

In this paper we use a purely hadronic scenario in our simulations. 
This means a very simple EoS \cite{gasEOS}, constructed under the
requirement that it 
reproduces saturation of the cold nuclear matter. 
This EoS is a
natural reference point for any other more elaborate EoS. 
We have found out that the original friction between baryon-rich fluids, 
incorporated in the model accordingly to estimate of Ref. \cite{Sat90}, is 
evidently insufficient to reproduce observable stopping of nuclear matter. 
Therefore, the original friction was enhanced 
by means of the tuning factor $\xi_h$, cf. Eq. (\ref{xi}), which is
plotted in Fig. \ref{fig15}. 
As seen, the enhancement factor turned out to be
unexpectedly large. However, it complies with earlier results of the  
2-fluid model with mean mesonic fields \cite{gsi94}. There it was
found out that  for the proper reproduction of the data on heavy-ion
collisions in the energy range  
from 0.4$A$ to 0.8$A$ GeV, where the purely hadronic scenario is
certainly applicable, the enhancement by a factor of 3 was required. 
From this large enhancement we infer that the rough estimate of
the  friction [24] is really too rough and that further more elaborate
microscopic calculations of the friction are required.
In doing this at lower energies, we have to rely on pure
hadronic effects, e.g. on the concept of the hadron liquid \cite{V04}
rather than hadron gas.
At higher incident energies the friction
enhancement could be 
associated with an indirect manifestation of the onset of the phase transition,
although the used EoS does not involve it directly. 
In Refs. \cite{SZ03,BLR04,Cassing05} it was argued that
just above the critical temperature, the quark-gluon system is
still a strongly interacting matter rather than a gas of perturbative
partons. This strong interaction anomalously enhances 
interaction cross sections in this phase.

With this simple hadronic EoS we have succeeded to reasonably
reproduce a great 
body of experimental data in the incident energy range
$E_{\scr{lab}}\simeq$ (1--160)$A$ GeV. The list includes proton
and pion rapidity distributions, proton transverse-mass spectra,
rapidity distributions of $\Lambda$ and  $\bar{\Lambda}$ hyperons and
antiprotons, elliptic flow of protons and pions (with the exception of
proton $v_2$ at 40$A$ GeV), multiplicities of pions, $\Lambda$ and
$\bar{\Lambda}$ 
hyperons, $K^+$ mesons and $\phi$ mesons, $\Xi^\mp$ and $\Omega^\mp$
multistrange  hyperons. 
However, we have also found
out certain problems. The calculated yield of $K^-$ is approximately
a factor of 1.5 higher than that in the experiment. 
We have also failed to describe directed transverse
flow of protons and pion at $E_{\scr{lab}}\geq 40A$ GeV. This
especially concerns pions, $v_1$ of which experimentally
anticorrelate with that of protons, while it correlates in our
calculations. This failure apparently indicates that the used EoS is
too hard and thereby leaves room for a phase transition, which always makes
the EoS softer. 

The used hadronic EoS allows modifications relating to different
incompressibilities ($K$) of the nuclear matter. We performed
simulations with soft ($K=$ 210 and 235 MeV) and hard ($K=$ 380 MeV)
EoS's. We have found out that the data certainly prefer soft EoS. 
All the calculations presented in this paper correspond to $K=$ 210
MeV.

The analysis of the global evolution of central
Pb+Pb collisions with present hadronic scenario shows in particular 
that incident energies (10--40)$A$ GeV, planned at the new GSI
facility, are favorable for production of
equilibrated matter with baryon densities 6 times higher than the
normal nuclear density. 
This conclusion is heavily based on the simple hadronic EoS used in
our calculations. However, we
expect that it is not too far from the truth, because we fairly well
reproduced the observed stopping power, which is of prime relevance to
the achieved compression.  
Moreover, in this energy range 
dynamical ($T,\mu_B$) trajectories of nuclear collisions pass in the
vicinity of the 
critical end-point, as it is estimated in Ref. \cite{Fodor01}. 
Note that the presently used EoS does not involve any phase
transition. In view of the latter, 
it is intriguing that discrepancies of the present hadronic scenario
with the data (i.e. in the form of the $\Lambda$ rapidity
distribution, proton and pion $v_1$ and $v_2$, and strangeness
production) are most
clearly seen at incident energies about 40$A$ GeV.

Our present experience shows that in order to conclude on the
relevance of a particular EoS the whole available set of data should
be analyzed in a wide incident energy range 
with the same fixed parameters of the model. 
This is one of the main  conclusions of our paper. 
Indeed, any
particular piece of data  can be apparently fitted by means of fine
tuning of these parameters, while to fit the whole set of them, one
needs a really good EoS. In the present paper we analyzed various data
only fragmentarily, just to fit the parameters and to give an overview of 
the resulting predictions, since the framework of a single paper does
not allow us 
to do more. Our preliminary conclusion is that the considered hadronic
EoS is not perfect at higher incident energies  $E_{\scr{lab}}\gsim
20A$ GeV because of the above mentioned problems. Detailed
analysis of  particular pieces of available data will be reported
in forthcoming papers. We are also going to extend our analysis to
other EoS's involving phase transition to the QGP.

\vspace*{5mm} {\bf Web page} \vspace*{5mm}

The source code of the 3-fluid model is publically available on the Web page 
http://theory.gsi.de/$\sim$mfd/. Free downloads of the code are available 
from there. Brief instructions on how to run this code are also given
on this page.

\vspace*{5mm} {\bf Acknowledgements} \vspace*{5mm}

We are grateful to E.L. Bratkovskaya, B.~Friman, M.~Gadzicki,
G.L.~Melkumov, I.N. Mishustin, J. Randrup, 
L.M. Satarov, V.V.~Skokov  and D.N. Voskresensky for fruitful
discussions. 
We are especially grateful to D.H. Rischke for careful reading of the
manuscript and making numerous useful comments. 
This work was supported in part by
the Deutsche  
Forschungsgemeinschaft (DFG project 436 RUS 113/558/0-3), the
Russian Foundation for Basic Research (RFBR grant 06-02-04001 NNIO\_a),
Russian Minpromnauki (grant NS-1885.2003.2).

\appendix

\section{Numerics}
\label{Numerics}

The relativistic 3D code for the above described 3-fluid model was
constructed by means of modifying the existing 2-fluid 3D code of
Refs. \cite{MRS88,MRS91}. The numeric scheme of the code is based
on the modified particle-in-cell method \cite{Roshal81,FCT-PIC},
which is an extension of the scheme first applied in Los-Alamos
\cite{Harlow76}. Taking advantage of modern computers, we have
comprehensively tested this numeric scheme in order to find
optimal parameters of the calculation.

In the particle-in-cell method the matter is represented by an
ensemble of  Lagrangian test particles. They are used for
calculation of the drift transfer of the baryonic charge, energy
and momentum. 
In the present scheme the test particle has a size of
the cell. Therefore, when a single test particle is moved on the grid,
it changes quantities in 8 cells, with which it overlaps. These
spatially extended particles make the scheme smoother and hence more
stable. 
The transfer due to pressure gradients,
friction between fluids and production of the fireball fluid is
computed on the fixed grid (so called Euler step of the scheme).
Simulation is performed in the frame of equal velocities of
colliding nuclei. In view of this combined nature of the numerical
scheme there are parameters of the test particles and the grid,
which are summarized in the Table \ref{tab:1}.

The basic scale of the numerical scheme is chosen along the $x$
direction (the transverse direction in the reaction plane).
According to the preset number of cells in the $x$ direction,
$L_x$, the maximal impact parameter $b_{\scr{max}}$, which is
foreseen for the simulation, and number cells free of the
matter, reserved for transverse expansion, $L_x^{\scr{free}}$, the
$\Delta x$ step of the grid is determined as follows
$$ \Delta x=\max\left\{2R_t;
R_p+R_t+b_{\scr{max}}\right\}/(L_x-L_x^{\scr{free}}), $$ where $R_p$ and
$R_t$ are radii of the projectile and target nuclei,
respectively\footnote{Here, we assumed that, if colliding nuclei
are different, then the larger one is target.}. Another transverse
step $\Delta y$ (in the out-of-reaction-plane direction) is taken
equal to $\Delta x$: $\Delta y = \Delta x$. $\Delta z$ and $\Delta
t$ are determined by means of preset ratios $\Delta x /\Delta t$
and $\Delta x /\Delta z$. The preset numbers of cells in the $y$
and $z$ directions ($L_y$ and $L_z$) are used to finally define
the size of the grid. The simulation takes place only in the
semi-space $y>0$ because of the symmetry of the system with respect
to the reaction plane. The number of baryon-rich test particles
($N_{\scr{tot}}^{\scr{bar}}=N_{\scr{projectile}}^{\scr{bar}}+
N_{\scr{target}}^{\scr{bar}}$) is specified at the
initialization step of the simulation. As for produced baryon-free
fluid, the actual total number of corresponding test particles
depends on the incident energy, the size of the cell and the  number
of produced fireball 
test particles per cell and per time step ($N_f$) and is
determined only upon the completion of the simulation.
The maximal total number of baryon-free test
particles $N_{\scr{tot}}^{\scr{fire}}$ should be larger than the
actual total number. 
$N_{\scr{tot}}^{\scr{fire}}$ is used to reserve memory for baryon-free test
particles.

\begin{table}[uhtl]
\begin{ruledtabular}
  \begin{tabular}{|c|ccc|}
 $E_{\scr{lab}}$, $A$ GeV& 1 $\div$  10 & 
$20\div158^*$ & $158^{**}$\\ \hline
 $L_x$& 
150 & 
320 & 320\\ 
 $L_y$& 
60 & 
130 & 130\\ 
 $L_z$& 
240 & 
820 & 820\\ 
 $L_x^{\scr{free}}$& 
5 & 
43 & 43\\ 
 $\Delta x/\Delta t$& 
3.5 & 
3.5 & 4\\ 
 $\Delta x/\Delta z$& 
1 & 
1 & 1.2\\ 
 $N_{\scr{tot}}^{\scr{bar}}$& 
$2\cdot10^6$ 
& $15\cdot10^6$ & $15\cdot10^6$ \\ 
$N_f$& 
2 
& 2 & 2 \\ 
 $N_{\scr{tot}}^{\scr{fire}}$& 
$3\cdot10^6$ 
& $25\cdot10^6$ & $25\cdot10^6$ \\ 
  \end{tabular}
\caption{Basic parameters of
the numeric scheme used for simulations of Au+Au and Pb+Pb
collisions at various incident energies $E_{\scr{lab}}$. 
}
\label{tab:1}
\end{ruledtabular}
\end{table}


In order to get reasonable accuracy in the simulation, the
following basic requirements should be met:
\\
(i) The grid should be extended enough to prevent escaping the
matter beyond this region before it gets frozen out.
\\
(ii) The grid in the beam ($z$) direction should be fine enough
for a reasonable description of the matter of initially
Lorentz-contracted nuclei. From the practical point of view, it is
desirable to have more than 30 cells on the Lorentz-contracted
nuclear diameter.
\\
(iii) 
The well-known Courant-Friedrichs-Lewy criterion states that 
the ratios of the space-grid
steps to the time step (e.g., $\Delta x /\Delta t$)
should be larger than 1 in order to have a consistent
and stable algorithm for solving hyperbolic partial-differential
equations. 
To avoid numerical diffusion, this ratios should be
taken optimal. As it was found in 1-dimensional simulations of
exactly solvable problems \cite{FCT-PIC}, the optimal range of
these ratios is $2.5<\Delta x /\Delta t<6$ with the preferable
$\Delta x /\Delta t \simeq 3.5$, minimizing the numerical
diffusion. 
This fact dictates the choice of equal-step grid in all
directions ($\Delta x : \Delta y :\Delta z = 1:1:1$), in spite of
Lorentz-contraction of incident nuclei, which is quite strong at
high energies. This choice makes the scheme isotropic with respect
to the numerical diffusion. However, in view of (i) and (ii)
requirements it makes the grid too fine in the transverse
directions and thus results in high memory consumption.
The need of the equal-step grid in all
directions for relativistic hydrodynamic computations within
conventional 1-fluid model was pointed out
in Ref. \cite{Rischke92}. As it was demontrated there, the matter
transport becomes even acausal if this condition is strongly violated. 
\\
(iv) The number of test particles per cell should be high enough
to avoid large numerical fluctuations in drift transfer. In
practice, this means that this number should be not less than 3
during the whole evolution of expanding matter, till its complete
freeze-out.

The final check of these requirements is possible only upon
completion of the simulation. The Table \ref{tab:1} presents the optimized set
of parameters for calculations of Au+Au and Pb+Pb collisions
at various incident energies. This set was determined by multiple
test runs of the code. 
The reaction Pb + Pb at $E_{\scr{lab}} = 158A$ GeV
is the most memory
and time consuming case considered in simulations. 
To completely fulfill
the requirement (iii) in this calculation we need 8 GB of memory for
the central collision (see column $20\div158^*$ in 
Table \ref{tab:1}).
In order to comply with memory restrictions, we still had to
take slightly unequal steps in different directions: 
$\Delta x : \Delta y :\Delta z = 1.2:1.2:1$, $\Delta x /\Delta t
= \Delta y /\Delta t = 4$ and $\Delta z /\Delta t = 3.3$  (see
column $158^{**}$  in Table \ref{tab:1}), 
for noncentral Pb + Pb 
collisions  at $E_{\scr{lab}} = 158A$ GeV.


The freeze-out criterion (\ref{FOcriterion}) is checked at the
Lagrangian step of the scheme. It is checked in the cells, which
overlap with the considered test particle. 
As mentioned above, the test particle has a
volume of the cell. Therefore, each test particle overlap with 8
cells. 
If the freeze-out criterion is met in {\em all
these cells}, this test particle is frozen out and removed from
further hydrodynamic evolution. Thus, the frozen-out test
particles are those droplets mentioned in Eq. (\ref{FOspectrum})
for the spectrum of observable hadrons. 
Precisely due to this
extended checking in 8 cells, the actual freeze-out energy
density\footnote{ 
related to the cell, where the center coordinate of the frozen-out
test particle belongs to
}
$\varepsilon_{\scr{frz}}$ 
participating in the freeze-out criterion (\ref{FOcriterion})
differs from that used in the code 
$\varepsilon_{\scr{frz}}^{\scr{code}}$. 
According to our estimate, the actual freeze-out energy density
$\varepsilon_{\scr{frz}}$ is approximately
twice as lower than $\varepsilon_{\scr{frz}}^{\scr{code}}$. 
This estimate was checked numerically. 
The reason behind this extended checking in 8 cells is
as follows. Had we checked the criterion only in a single cell,
where the test particle is located, we would prevent any
hydrodynamic expansion of the system. Indeed, if the matter,
during its expansion, starts to fill in a cell, which was empty
before that, it occupies less than 1/3 of this cell during the
single time step, because the time step should be less than the
space step (cf. $\Delta x /\Delta t = 3.5$) for this numerical
scheme. Then, all the densities in this newly filled-in cell are
quite low at this time step. Therefore, the freeze-out criterion
is certainly fulfilled in this cell. Had we confined ourselves to
checking only this cell, we would immediately freeze the matter in
this cell out, thus preventing hydrodynamic expansion of the
matter.

In order to avoid formation of bubbles of frozen out matter inside the 
dense environment, we introduced additional criterion of the freeze-out: 
at least one of the above discussed cells should be empty, i.e. contain 
no test particles\footnote{
By definition, the cell is empty, if 
no center coordinate of any test particle belongs to this cell.
}.
This additional criterion means that the analyzed 
test particle is located near the surface of the system, provided, of course, 
that the number of test particles per cell is large enough 
(cf. condition (iv)), which excludes origination of such empty cells due to 
fluctuations.

\begin{widetext}
\section{Resonance Decays}
\label{Resonance Decays}
\subsection{General Consideration}

Contribution of resonance decays into stable particle spectra has been
studied long ago \cite{Kapusta77} and later analyzed in more detail
\cite{Sollfrank91}. To make the paper self-contained, we briefly
present here formulas for these decays,  which in fact are the same as
those in \cite{Kapusta77,Sollfrank91}  but differ from those in
presentation, which, from our point of view,  is advantageous for
numerical realizations.  

Consider a hadronic resonance $R$ with degeneracy factor $g_R$ and a
spectral function $A_R(s)$,  where $s$ is the resonance mass squared,
which decays into $n$ particles  
$$ R \to 1+2+...+n$$ of masses $m_1$, $m_2$,..., $m_n$, through a $J$th
channel of its decay with a branching  ratio $b_J$. Then the
distribution of the "1" particle, resulting from this decay, is as
follows   
\begin{eqnarray}
\label{RJ-to-n}
&&\left(E_1\frac{d^3 N_1^{(R \to 1+2+...+n)}}{d^3 p_1}\right)_J = 
\cr
&=& b_J
\int_{\left(\sum_{i=1}^n m_i\right)^2}^\infty ds \; A_R(s)
\int \frac{d^3 p_R}{E_R}
\left(E_R\frac{d^3 N_R (s,p_R)}{d^3 p_R}\right)
\left(E_1\frac{d^3 \Gamma_1^{(R \to 1+2+...+n)} (s,p_R,p_1)}{d^3 p_1}\right). 
\end{eqnarray}
Here, the distribution of the $R$ resonance, produced by the
hydrodynamic computation,  is  
\begin{eqnarray}
\label{R-dist}
E_R\frac{d^3 N_R (s,p_R)}{d^3 p_R}= 
\frac{g_R V}{(2\pi)^3}
\frac{u\cdot p_R}{\exp[\beta(u\cdot p_R-\mu_R)] \pm 1}, 
\end{eqnarray}
where $V$ is a small proper (i.e. in the rest frame) volume of the
fluid element,  $g_R$ is degeneracy of the $R$ resonance, 
$u$ is the hydrodynamic 4-velocity, $\beta$ is the
inverse temperature, $\mu_R$  is the chemical potential, $p_R$ is the
4-momentum of the $R$ resonance,  $s=p_R^2$, the upper (lower) sign in
this expression corresponds to baryonic  (mesonic) resonances. The
distribution of the "1" particle, resulting  from decay of a single
$R$ resonance, is expressed as follows 
\begin{eqnarray}
\label{G-to-n}
E_1\frac{d^3 \Gamma_1^{(R \to 1+2+...+n)} (s,p_R,p_1)}{d^3 p_1}= 
\frac{1}{2R_n(s)} \int 
\left(\prod_{i=2}^n\frac{d^3 p_i}{2E_i}\right)
\delta^4 \left(p_R - \sum_{i=1}^n p_i\right)
\end{eqnarray}
under assumption that the matrix element of the decay is constant all
over the available $n$-particle phase-space volume \cite{Byckling},
which, in its turn, reads 
\begin{eqnarray}
\label{Rn}
R_n(s)= \int 
\left(\prod_{i=1}^n\frac{d^3 p_i}{2E_i}\right)
\delta^4 \left(p_R - \sum_{i=1}^n p_i\right). 
\end{eqnarray}
In order to obtain the total contribution of the $R$ decay into "1"
particle spectrum, we should sum over all $J$ branches of the $R$
decay, in which the "1" particle appears, and take into account that
particles identical to "1" can also be among residue particles
$2,3,...,n$ 
\begin{eqnarray}
\label{R-to-n}
E_1\frac{d^3 N_1^{(R \to 1+X)}}{d^3 p_1} =
\sum_J N_1^J 
\left(E_1\frac{d^3 N_1^{(R \to 1+2+...+n)}}{d^3 p_1}\right)_J. 
\end{eqnarray}
Here $N_1^J$ is the multiplicity of "1" particles in the $J$th channel
of the $R$ decay. Here and below, $E_R=\left(s + {\bf
    p}_R^2\right)^{1/2}$ and  
$E_i=\left(m_i^2 + {\bf p}_i^2\right)^{1/2}$.  

Fortunately, the the major part of integrations involved in
Eqs. (\ref{RJ-to-n}) and (\ref{G-to-n}) can be performed
analytically. We explicitly present results only for two-particle and
three-particle decays, which are dominant for resonances under
consideration.      

\subsection{Two-Particle Decays}

The result for a two-particle decay of a $R$ resonance is as follows 
\begin{eqnarray}
\label{RJ-to-2}
&&\left(E_1\frac{d^3 N_1^{(R \to 1+2)}}{d^3 p_1}\right)_J 
\cr
&=& b_J
\frac{g_R V}{(4\pi)^2}
\frac{1}{\left[(u\cdot p_1)^2-m_1^2\right]^{1/2}}
\int_{\left(m_1+m_2\right)^2}^\infty ds \; \frac{A_R(s)}{R_2(s)}
\sum_{n=1}^\infty (\mp 1)^{n-1} 
\left[
\frac{n\beta E_R + 1}{(n\beta)^2} \exp\left\{-n\beta(E_R-\mu_R)\right\}
\right]_{E^+_R}^{E^-_R}, 
\end{eqnarray}
where the upper (lower) sign in this expression corresponds to
baryonic (mesonic) resonances,  
\begin{eqnarray}
\label{R2}
R_2(s)=\frac{\pi}{2s} \lambda^{1/2} (s,m_1^2,m_2^2)
\end{eqnarray}
is the 2-particle phase-space volume and 
\begin{eqnarray}
\label{ER}
E^\pm_R= 
\frac{1}{2m_1^2} 
\left[\left(s+m_1^2-m_2^2\right)\left(u\cdot p_1\right)
\pm
\lambda^{1/2} (s,m_1^2,m_2^2)\left[(u\cdot p_1)^2-m_1^2\right]^{1/2}
\right]. 
\end{eqnarray}
Here and below 
\begin{eqnarray}
\label{lambda}
\lambda (x,y,z)= (x-y-z)^2 -4yz
\end{eqnarray}
is the standard kinematic function \cite{Byckling}. 
To arrive to this result, the following expansion 
\begin{eqnarray}
\label{expan}
\frac{1}{\exp[\beta(u\cdot p_R-\mu_R)] \pm 1}= 
\sum_{n=1}^\infty (\mp 1)^{n-1} 
 \exp\left\{-n\beta(u\cdot p_R-\mu_R)\right\}
\end{eqnarray}
was used, the advantage of which is that it is rapidly convergent, when 
$u\cdot p_R > \mu_R$. The opposite limit, i.e. $u\cdot p_R < \mu_R$,
corresponds to low  temperatures and hence hardly contributes to the
resonance production.  The $n=1$ term in this expansion, as well in
Eq. (\ref{RJ-to-2}), corresponds to the classical J\"uttner distribution.

\subsection{Three-Particle Decays}

The result for a three-particle decay of a $R$ resonance is as follows 
\begin{eqnarray}
\label{RJ-to-3}
\left(E_1\frac{d^3 N_1^{(R \to 1+2+3)}}{d^3 p_1}\right)_J 
&=&
b_J
\frac{g_R V}{32\pi}
\frac{1}{\left[(u\cdot p_1)^2-m_1^2\right]^{1/2}}
\int_{\left(m_1+m_2+m_3\right)^2}^\infty ds \; \frac{A_R(s)}{R_3(s)}
\cr
&\times&
\int_{\sqrt{s}}^\infty dE_R 
\theta\left(X^+/X^-\right)
\cr
&\times&
\theta\left(\frac{\left(m_2^2-m_3^2\right)^2-\left(m_2^2+m_3^2\right)X^+ +\left|m_2^2-m_3^2\right|\sqrt{\lambda^+}}%
{\left(m_2^2-m_3^2\right)^2-\left(m_2^2+m_3^2\right)X^- +\left|m_2^2-m_3^2\right|\sqrt{\lambda^-}}
\right)
\theta\left(\frac{-\left(m_2^2+m_3^2\right)+X^+ +\sqrt{\lambda^+}}%
{-\left(m_2^2+m_3^2\right)+X^- +\sqrt{\lambda^-}}
\right)
\cr
&\times&
\left[
\left(\sqrt{\lambda^+}-\sqrt{\lambda^-}\right)
+
\left|m_2^2-m_3^2\right|\ln\frac{X^+}{X^-}
\right.
\cr
&-&
\left|m_2^2-m_3^2\right|\ln
\frac{\left(m_2^2-m_3^2\right)^2-\left(m_2^2+m_3^2\right)X^+ +\left|m_2^2-m_3^2\right|\sqrt{\lambda^+}}%
{\left(m_2^2-m_3^2\right)^2-\left(m_2^2+m_3^2\right)X^- +\left|m_2^2-m_3^2\right|\sqrt{\lambda^-}}
\cr
&-&
\left.
\left(m_2^2+m_3^2\right)\ln
\frac{-\left(m_2^2+m_3^2\right)+X^+ +\sqrt{\lambda^+}}%
{-\left(m_2^2+m_3^2\right)+X^- +\sqrt{\lambda^-}}
\right]
\frac{E_R}{\exp\left[\beta\left(E_R-\mu_R\right)\right] \pm 1}
\end{eqnarray}
where the upper (lower) sign in this expression again corresponds to baryonic (mesonic) resonances, 
\begin{eqnarray}
\label{Xpm}
X^+ &=& \min \left(\widetilde{X}^+,X_{max}\right), 
\quad
X^- = \max \left(\widetilde{X}^-,X_{min}\right), 
\\
\label{Xmm}
X_{max} &=& \left(s^{1/2}-m_1\right)^2, 
\quad
X_{min} = \left(m_2+m_3\right)^2, 
\\
\label{tXpm}
\widetilde{X}^\pm&=& 
s+m_1^2-2E_R\left(u\cdot p_1\right)\pm
2\left(E_R^2-s\right)^{1/2}\left[(u\cdot p_1)^2-m_1^2\right]^{1/2}, 
\\
\label{Lpm}
\lambda^\pm &=& \lambda \left(X^\pm,m_2^2,m_3^2\right),
\\
\label{R3}
R_3(s)&=& \left(\frac{\pi}{2}\right)^2 \frac{1}{s}
\int_{X_{min}}^{X_{max}} 
\frac{dX}{X} 
\lambda^{1/2}\left(s,X,m_1^2\right)
\lambda^{1/2}\left(X,m_2^2,m_3^2\right),  
\end{eqnarray}
$\theta(\lambda^+)$, $\theta(\lambda^-)$, etc. are step functions which
define the accessible kinematic region. 
\end{widetext}

\end{document}